WORKING PAPER

# On Labs and Fabs: Mapping How Alliances, Acquisitions, and Antitrust are Shaping the Frontier AI Industry

Tomás Aguirre[1]

June, 2024

*latest version available here;*

*mapping available here*

**Abstract**

As frontier AI models advance, policy proposals for safe AI development are gaining increasing attention from researchers and policymakers. This paper explores the current integration in the AI supply chain, focusing on vertical relationships and strategic partnerships among AI labs, cloud providers, chip manufacturers, and lithography companies. It aims to lay the groundwork for a deeper understanding of the implications of various governance interventions, including antitrust measures. The study has two main contributions. First, it profiles 25 leading companies in the AI supply chain, analyzing 300 relationships and noting 80 significant mergers and acquisitions along with 40 antitrust cases. Second, we discuss potential market definitions and the integration drivers based on the observed trends. The analysis reveals predominant horizontal integration through natural growth rather than acquisitions and notable trends of backward vertical integration in the semiconductor supply chain. Strategic partnerships are also significant downstream, especially between AI companies and cloud providers, with large tech companies often pursuing conglomerate integration by acquiring specialized AI startups or forming alliances with frontier AI labs. To further understand the strategic partnerships in the industry, we provide three brief case studies featuring companies like OpenAI and Nvidia. We conclude by posing open research questions on market dynamics and possible governance interventions, such as licensing and safety audits.

[1] University of São Paulo and Palver. This is a preliminary version of the paper. Comments are welcome. E-mail: t6aguirre@gmail.com

# Executive summary

**Background and objective:** The AI supply chain is complex and rapidly evolving, presenting challenges similar to those faced during the rise of the internet and the digital economy. Barriers to entry, such as low marginal costs and potential network effects, can grant companies significant market power. It remains unclear whether current economic theories can fully explain these dynamics or if new models are needed. This paper examines the current state of integration within the leading AI supply chain, focusing on vertical integration and strategic partnerships. Our goal is to map the AI supply chain comprehensively and consider the implications for regulatory proposals and the impact of antitrust laws.

**Methodology:** Through a broad search, we investigated key companies in this supply chain, including AI labs and chip designers. Our analysis encompasses:

- Examining connections between 25 major companies and their pairwise relationships, observing a widespread presence of strategic partnerships.
- Investigating three specific cases to better understand these alliances.
- Listing about 80 major mergers and acquisitions over $100 million and pointing out 40 relevant antitrust court cases.

**Key Findings:**

- Horizontal integration throughout the AI supply chain is mainly driven by market consolidation through natural growth, not acquisitions
- There is a noticeable trend of backward vertical integration in both the lithography and the chip manufacturing industries.
- Downstream in the chain, we predominantly have a significant amount of strategic partnerships between AI labs and cloud companies.
- Big tech companies frequently make conglomerate integration by buying startups on narrow AI applications or setting up strategic partnerships with foundation AI labs.
- The three case studies that we conducted are illustrative of trends in the AI supply chain:

- ○ **OpenAI and Microsoft**: The partnership between Azure and OpenAI led to the development of a top 5 supercomputer in which GPT-3 was trained. This partnership is an illustration of how AI labs try to secure access to compute at scale by partnering with big cloud providers while big tech companies try to incorporate AI in their wide portfolio of
- ○ **ASML, TSMC, Samsung, and Intel**: the partnership was essential to the development of EUV technology, in which the development of frontier AI accelerators relies upon.
- ○ **Nvidia and Arm**: this was the first litigated major vertical integration attempt that was terminated in the AI supply chain. The main argument of the FTC was that Nvidia would be able to conduct anti-competitive behavior by foreclosing the access of ARM's Core IP to its competitors. This may indicate a trend for antitrust authorities to pay increased attention to the semiconductor market.

- Governmental actions significantly shape the AI supply chain through subsidies, sanctions, and industrial policy. However, there has been limited antitrust action on the supply chain, with the most noteworthy cases being Applied Materials and Tokyo Electron, and ARM and Nvidia.
- Various drivers may be behind these integrations, ranging from economic synergies and strategic competition to governmental interventions. We tentatively conclude that what is leading to scenario with that much quasi-vertical integration is i) companies seeking to ensure compute access for doing large training runs, ii) big tech companies balancing specialization with broad capabilities: iii) high transaction costs in R&D, specially for companies upstream that develop or deeply engage with EUV technology for chip manufacturing, iv) a market in its initial stages of development and hence which has not developed which large markets for doing do major, impersonal transactions since there are no established ways of working. iv) desire for companies to be secretive. Finally, we believe that these factors are boosted by an emergent "winner-takes-all" sentiment within the industry. We have seen indications of these hypotheses as we lay out in the report, but such hypotheses still need further analysis and empirical testing before we can assert with confidence.

**Key open questions:**

1. **How does the prevailing market structure influence the trajectory of AI industry advancements?** Regulatory effectiveness for AI is likely to be influenced by the AI industry's structure, particularly in how it manages vertical relationships. Regulatory implications could involve trade-offs between competition and safety. For instance, slowing AI development to address risks might conflict with antitrust policies aimed at promoting competition. There may also be significant tension between enhancing short-term consumer welfare and mitigating long-term risks from advanced AI systems. National security concerns further complicate the picture, as exemplified by the Qualcomm-Broadcom merger block, where market power and security interests were at odds.
2. **How does the current market structure within the AI supply chain affect the implementation and effectiveness of current regulatory proposals?** Vertical integration's impact on transparency and regulatory compliance seems to be two-fold: it could hinder the enforcement of rules requiring the reporting of key inputs such as compute usage due to a lack of public information, yet it might also better position companies to comply with strict privacy and cybersecurity standards. Moreover, a more concentrated AI supply chain could facilitate coordinated efforts for industry standard-setting, which could be beneficial for governance, although it could also raise antitrust concerns over potential collusion.
3. **Will structural remedies be necessary to establish effective regulatory frameworks in the AI industry?** The necessity for structural remedies in regulatory frameworks could arise from the varying consequences of different types of market integration and policy goals. For instance, to slow AI advancement, one might advocate for a more horizontally integrated market to reduce competitive race dynamics. However, this approach also raises concerns about power concentration. Unbundling practices, similar to those in the electricity and railway sectors, may also be relevant, particularly for third-party reporting mechanisms in AI.

All these questions would benefit from empirical groundwork that, for instance, estimates the production function of chip fabricators, assesses the market demand elasticity of AI accelerators, or tests which drivers for integration are most relevant. Research on how vertical integration has impacted the effectiveness of regulations in other industries—especially dual-use or general-purpose technologies—would be highly valuable. This paper provides an initial exploration of how different kinds of integration might affect regulatory strategies like licensing and standard-setting, identifying areas where further research is needed.

**Further Details:** We provide a comprehensive mapping of the industry on the link. Additional information is available in the appendices, covering topics such as the mapping of the AI supply chain, and the brief case studies on strategic partnerships and alliances, featuring companies like OpenAI, Microsoft, ASML, Nvidia, and ARM.

## Acknowledgments

I am grateful to Andrew J. Koh and Charlotte Siegmann for their invaluable mentorship. I acknowledge the Swiss Existential Risk Initiative and the Cambridge-Boston Alignment Initiative for funding and support. Special thanks to Michael Aird, Renan Araujo, Tobias Häberli, Andreas Haupt, Shin-Shin Hua, Leonie Koessler, Claudio Lucinda, Luís Mota Freitas, and Konstantin Pilz for their comments. I also appreciate contributions from anonymous experts. Any remaining errors are my own.

# 1. Introduction and Motivation

The development of artificial intelligence into a general-purpose technology this century could be as transformative as the internet or even the Industrial Revolution (see, e.g., Trammell and Korinek, 2023; Erdil and Besiroglu, 2023; Goldfarb et al, 2023; Elondou et al, 2023). Along with great opportunities, advanced AI systems also pose various risks, including cybersecurity threats, biological vulnerabilities, potential for social manipulation, worsening economic inequality, and reinforcing societal biases (see, e.g., Bengio et al, 2023; Acemoglu, 2021; Brundage et al., 2018). To address these challenges, most AI policy experts support practices such as pre-deployment risk assessments, third-party model audits, and safety restrictions (Schuett et al., 2023). . Concerns about AI risks have already led to several responses, including President Biden's Executive Order on AI, the European Union's AI Act, the United Kingdom's AI Safety Summit, and the United Nations Secretary-General's creation of an advisory body dedicated to AI governance.

As noted by Cullen (2020) and Belfield and Hua (2022), antitrust considerations may affect or complement regulatory proposals for frontier AI models. This paper aims to provide a comprehensive overview of the current AI supply chain, focusing on companies that may be critical in developing transformative AI systems. Special emphasis is placed on vertical integration and strategic alliances between these companies. We will focus primarily on foundation models, defined as AI models that are "trained on broad data at scale and are adaptable to a wide range of downstream tasks" (Bommasani et al., 2021).

Throughout this report, we will focus on the supply chain required to train large foundation models with general capabilities ranging between GPT-3.5 and GPT-4, which we classify as frontier. We categorize models with capabilities comparable to GPT-3 to GPT-3.5 as non-frontier. Our analysis primarily considers products perceived by consumers as rough substitutes, as we believe this dimension will be the most important one for both regulatory and antitrust interventions. Tentative relevant market definitions are established for each supply chain step, including lithography companies, cloud providers and chip designers.

This paper focuses primarily on the compute used for the training and deployment of AI models. We have chosen this focus for two main reasons: first, because alongside algorithms and data, compute is one of the three most significant inputs in AI development; and second, because unlike data or algorithms, compute is easily measurable and is a rival and excludable asset, making it easier to implement effective oversight mechanisms.

We profiled 25 leading companies in the AI industry and mapped 300 pairs of relationships between these companies and identified approximately 80 actions and mergers worth more than USD 100 million in which these companies were involved. Additionally, we documented other relevant events, such as major investments and disinvestments. We mapped approximately 50 antitrust cases and conducted three brief case studies on the industry, including the OpenAI and Microsoft partnership and ASML's partnership with Intel, Samsung, and TSMC. Furthermore, we discussed some potential drivers that could explain why integration in the industry might be occurring. The comprehensive mapping is available in the appendices as well as an online resource. The paper concludes by listing open questions and pointing out gaps in the existing literature accompanied by a preliminary discussion about how to best think about the market structure of different steps of the AI supply chain.

We also identify important open questions that need further exploration by industrial economists, competition lawyers, and regulatory authorities. To highlight potentially fruitful research areas, we start by discussing the market structure of various segments within the AI supply chain and the related trade-offs. Understanding this industry presents significant challenges, similar to the economic puzzles that emerged with the rise of big tech platforms or the development of open-source software in the early 2000s.

| Name | Lithography Companies | AI Chip Fabricators | AI Chip Designers | B2B Cloud | AI Lab |
|---|---|---|---|---|---|
| Alphabet | No | No | Yes, frontier | Yes, frontier | Yes, frontier |
| Amazon | No | No | Yes, non-frontier | Yes, frontier | Yes, non-frontier |
| AMD | No | No | Yes, non-frontier | No | No |
| Anthropic | No | No | No | No | Yes, frontier |
| Apple | No | No | Yes, non-frontier | No | Yes, non-frontier |
| ASML | Yes, frontier | No | Yes, non-frontier | No | No |
| Broadcom | No | No | Yes, non-frontier | No | No |
| Canon | Yes, non-frontier | No | No | No | No |
| Cerebras | No | No | Yes, non-frontier | No | No |
| Cohere | No | No | No | No | Yes, non-frontier |
| GlobalFoundries | No | Yes, non-frontier | No | No | No |
| Hugging Face | No | No | No | No | Yes, non-frontier |
| Inflection AI | No | No | No | No | Yes, frontier |
| Intel | No | Yes, non-frontier | Yes, non-frontier | Yes, non-frontier | No |
| Meta (Facebook) | No | No | Yes, non-frontier | No | Yes, frontier |
| Microsoft | No | No | Yes, non-frontier | Yes, frontier | Yes, frontier |
| Mistral | No | No | No | No | Yes, non-frontier |
| Nikon | Yes, non-frontier | No | No | No | No |
| Nvidia | No | No | Yes, frontier | No | Yes, non-frontier |
| OpenAI | No | No | No | No | Yes, frontier |
| Oracle | No | No | No | Yes, non-frontier | No |
| Samsung | No | Yes, non-frontier | Yes, non-frontier | No | Yes, non-frontier |
| Softbank (Arm) | No | No | Not exactly | No | No |
| TSMC | No | Yes, frontier | No | No | No |
| xAI | No | No | No | No | Yes, non-frontier |

*Table 1: Summary of vertical integration in the AI supply chain (snapshot from May 2024)*[2]

[2] Table created by the author. While we tried to assess based on both technical and market reports, it is important to note that the binary classification between frontier and non-frontier is a simplification of a dynamic scenario. See Section 3 for details and discussion on tentative definition of relevant markets.

## 1.1 Relevant literature

This paper engages with three main bodies of literature: regulatory frameworks for frontier AI models; competition policy for the technological sector; and the relationship between regulation and antitrust.

### 1.1.1 AI regulation

When a technology can do significantly good but also significantly harm, the optimal deployment rate may be slower, as society may learn about the risks during deployment (Acemoglu and Lensman, 2023). Deployment should potentially also be delayed until further investments in safety because, when welfare levels rise, the risks become more significant compared to the value of the technology (Jones, 2016; Jones, 2023).

Recently, regulatory proposals have increasingly focused on frontier AI models that require large training runs with AI accelerators — chips specifically designed for training or inference of machine learning models. These training runs are typically characterized by the substantial number of Floating Point Operations (FLOPs) used and are usually benchmarked against existing deployed products and empirical patterns of how performance increases with model size to predict their potential capabilities. For example, Biden's AI executive order defines "dual-use foundation models" as models that used more than 10^23 FLOPs if trained with biological sequence data or 10^26 FLOP otherwise (White House, 2023).

AI governance proposals include permitting, which requires actors to meet certain safety criteria to obtain licenses (see, e.g., Higgins, 2023); auditing, involving external reviews of AI systems for regulatory compliance (Mökander et al., 2023); liability regimes, which establish accountability in cases where AI causes harm (see, e.g., Llorca et al., 2023); information sharing and incident reporting (see, e.g., Stafford and Trager, 2022); compute taxation or subsidy mechanisms designed to incentivize resource allocation toward safety-conscious AI development (see, e.g., Jensen et al., 2023); and regulatory markets (Hadfield and Clark, 2023).

### 1.1.2 Competition policy in the technological sector

The AI supply chain includes one of society's most complicated technologies, is capital intensive and concentrated. For instance, ASML is the single provider of extreme-ultraviolet machines (EUV) lithography machines needed for AI chip production, TSMC and Samsung are the single companies capable of building the most advanced AI accelerators—specialized chips for AI training —, and NVIDIA is the single market supplier of frontier GPUs. There are also a lot of vertical integration and strategic partnerships that resemble vertical integration in the frontier AI labs. Microsoft has stakes in two of the most advanced AI labs—Inflection and OpenAI – (Silicon, 2023), while Alphabet is the owner of DeepMind and has a stake in Anthropic (Bloomberg, 2023). They are also major players in the cloud computing market and develop in-house advanced AI applications (Allied Market Research, 2023). Google, additionally, designs its own AI accelerators called Tensor Processing Units (Reuters, 2023), with Microsoft reportedly following this trend of creating its own chips for deep learning tasks (Reuters, 2023). Apple, Meta, and Amazon are also deeply involved in different steps of the AI supply chain (see, e.g., Rikap, 2023).

This indicates that, similar to the hardware and software industries (see, e.g., Shy, 2001; Tirole, 2023), high fixed costs, low marginal costs, network externalities, and product differentiation are prevalent, raising concerns about entrenched market power. However, there is yet substantial uncertainty about how to best understand the market structure and competition dynamics of different steps of the AI supply chain. Vipra and Korinek (2023) have argued that the cost development of foundation models and the associated necessary infrastructure make it resemble a natural monopoly, suggesting a path towards regulation as a utility like electricity and transit. However, as this is not a yet consolidated market that the role of, e.g., product differentiation and contestability will have, the scenario remains very uncertain.

In a roundtable on competition policy and generative AI, Acemoglu argued for antitrust laws to foster alternatives to dominant tech companies and reduce their social and economic power. Athey, the current Chief Economist of the Antitrust Division at the U.S. Department of Justice, emphasized the challenge's resemblance to the rise of multi-sided markets and platforms. Lina M. Khan, in her paper "Amazon's Antitrust Paradox" (2017), argued that conglomerate and vertical integration by big tech should be analyzed for their dynamic effects on market structure.

Since becoming chair of the FTC, Khan has challenged Microsoft's acquisition of Blizzard, Meta's acquisition of a VR startup, and has opened litigation against Google and Amazon. Her leadership marks a significant shift in U.S. antitrust policy, especially regarding the tech sector. The OECD has published a paper on "Theories of Harm for Digital Mergers," highlighting ecosystem-based and privacy-focused theories and incorporating long-term effects in competition policy. How these perspectives will shape competition authorities' approach to foundation model markets remains uncertain.

### 1.1.3 Interplay between regulation and antitrust

While AI safety considerations would probably fall outside the scope of antitrust enforcement, it is crucial to examine how competition policies could influence regulatory proposals in the AI industry. Market structure, for example, can impact the level of R&D investment in an industry (see, e.g., Armour and Teece, 1980), suggesting that antitrust policies can affect the rate at which frontier AI systems evolve. The impact of vertical integration on R&D is theoretically ambiguous, demanding empirical investigation into the specificities of the AI supply chain. Information sharing and incident reporting policies between companies in the industry may also raise collusion concerns by competition authorities.

Increased vertical integration might, by default, lead to less public information about the industry, strengthen the industry lobby, inflate profit margins, and result in greater power accumulation. Conversely, a highly vertically integrated market could potentially facilitate the diffusion of safety standards and enhance the capacity for a timely, coordinated response to risks arising from the training or deployment of foundation models. Additionally, the vertical relationships and contracts established between companies in a supply chain with oligopolistic aspects at each level of the chain, as extensively discussed by Lee et al (2021).

This points to the possible necessity of structural remedies in certain regulatory proposals. As suggested by Narechania and Sitaraman (2023), regulations could potentially mandate the separation of various business activities within a single AI firm, such as the design of foundation models and the ownership of the data centers where they are trained. As the relationship between industry integration and safety remains unclear, further research is needed on the topic.

## 1.2 Limitations

While this paper provides a comprehensive overview of vertical and horizontal integration within the AI supply chain, it has several limitations. Firstly, the rapidly evolving nature of AI technologies and policies can quickly outdated some of our findings. Secondly, the paper focuses on major players and high-value transactions, which may not capture the full diversity of the AI landscape, including smaller entities and emerging markets. Thirdly, we decided to exclude China from our analysis driven by challenges in accessing reliable data, and significant regulatory and policy differences. Fourthly, our study is mostly confined to available data and published research as well as conversations with experts anonymously, and may not reflect undisclosed strategic partnerships or unpublished technical developments.

# 2. Understanding the AI Supply Chain

This section will be an overview of the AI supply chain. For readers familiar with this, we recommend skipping to the next section.

## 2.1 Background history

In 1948 at Bell Telephone Laboratories, a team led by physicists John Bardeen, Walter Brattain, and William Shockley created the first transistor, a semiconductor device used to amplify or switch electronic signals. Until then, the electronics industry was dominated by more sizable and less energy-efficient vacuum tubes. This won them a Nobel Prize in Physics in 1956 and laid the foundation for increasingly powerful digital systems (Shaller, 1997).

In the same year of 1956, Dartmouth College held a conference that laid the foundation of artificial intelligence as a distinct academic discipline. Organized by John McCarthy, Marvin Minsky, Nathaniel Rochester, and Claude Shannon, the conference, which in its proposal stated that "every aspect of learning or any other feature of intelligence can in principle be so precisely described that a machine can be made to simulate it" (McCarthy et al, 1955). Although the initial optimism for AI's potential was later met with challenges and periods of skepticism known as "AI winters", over time the semiconductor industry and the field of artificial intelligence have developed a deeply interconnected relationship.

The evolution of the semiconductor industry has allowed AI models to go from simple rule-based systems to deep learning models with billions of parameters (LeCun, Yoshua & Bengio, 2015). In 2006, Microsoft researchers (Chellapilla et al 2006) recognized that convolution neural networks (CNN), designed in the 1990s to process images, could be trained more efficiently by parellezing the training using NVIDIA's Graphics Processing Units, first designed for video games. Developing on this idea, the AlexNet algorithm won the Stanford image-classification competition in 2012 (Krizhevsky et al, 2012). With the introduction of transformers in 2017 by Google Brain researchers, natural language processing tasks - which are sequential in nature - became more easily parallelized using hardware accelerators, allowing breakthroughs in machine learning.

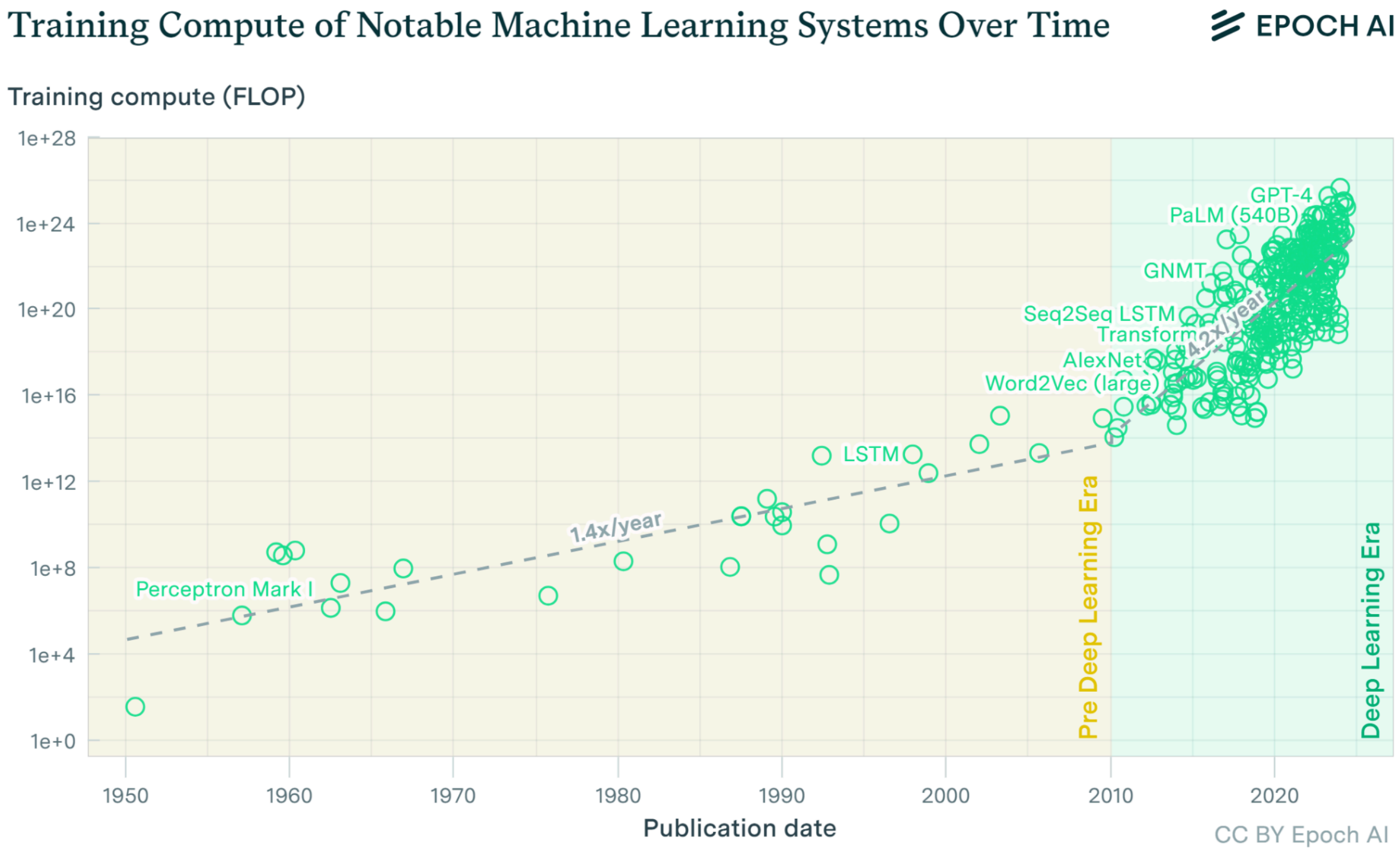

*Phases of AI development by amount of training compute (EpochAI, 2023)*

## 2.2 Inputs necessary for development of frontier AI models

State-of-the-art AI labs develop foundation models, which are large models capable of being applied across various applications (see, e.g., Bommasani et al., 2021). The development of these models requires three primary inputs: data, algorithms, and computing resources (see, e.g., Buchanan, 2020). The rise of deep learning since the early 2010s has been driven by the availability of large amounts of data, advances in neural network architectures, and substantial improvements in computer power - compute, henceforth.

### 2.2.1 Data

Vast amounts of data are necessary for training frontier AI models. AI labs commonly use large public datasets, like Common Crawl and Wikipedia, while often supplementing these with proprietary datasets, especially for niche applications. The quality of data is critical; for supervised learning approaches, datasets often need to be meticulously labeled to train the models effectively; other approaches such as self-supervised learning also commonly depend on human-labeling efforts at different stages. Over time, there has been an exponential increase in data availability, which has helped the advancement of more complex and accurate models (see, e.g., Villalobos et al., 2022).

### 2.2.2 Algorithms

Frontier AI models are based on different architectures of neural networks, designed to learn data patterns through the gradual optimization of model parameters to minimize loss functions. According to Erdil & Beriglu (2022), the algorithmic efficiency of neural networks doubles every 9 months, much faster than Moore's law. That means that current models can achieve similar performance levels to older ones with fewer compute and data. Both algorithms and data can be considered non-rival but excludable; that is, multiple users can utilize the same algorithms and data simultaneously, yet it is possible to prevent others from using them.

### 2.2.3 Compute

Computing resources (short: compute) are essential for both the training and deployment of AI models. These often include specialized hardware such as AI accelerators, which are chips optimized for AI computations. Graphics Processing Units (GPUs) are among the most widely used types of AI accelerators (Reuther et al., 2023). Field-Programmable Gate Arrays (FPGAs) are another type of accelerator, characterized by being able to be reprogrammed to suit specific computational tasks after manufacture. Application-specific integrated Circuits (ASICs), like Google's Tensor Processing Unities (TPUs), are designed for a specific function; According to a study by Sevilla et al (2022), the amount of compute required by frontier AI systems has increased by a factor of 4.2 every year since 2010. Unlike data and algorithms, computing resources are rivalrous: one person using a chip for a purpose directly impedes others from using it. This characteristic, along with the ease of measuring and tracking compute resources compared to the other inputs needed for AI development, positions it as a key element in AI governance proposals.

Additionally to the chips themselves, the operation of data centers and the infrastructure necessary to maintain them are of great importance in the overall functioning of the industry.

These are large-scale facilities that demand significant use of electricity and water. Data centers and cloud providers are where the training of large foundation models actually happens, usually utilizing the same facilities as other non-AI applications. As their size is pushed to the limit, it is becoming increasingly important to handle how these chips are put together and kept at adequate temperatures (Pilz and Heim, 2023).

### 2.2.4 Scaling laws

AI models generally exhibit a strong relationship between performance on their training objectives and factors such as model size, data, and compute usage (Clark et al., 2022). Kaplan et al. (2020) note that "performance has a power-law relationship with each of the three scale factors N [number of parameters], D [dataset size], C [compute utilized]," while it "depends […] weakly on model shape." Moreover, the relationship between pretraining compute budget and held-out loss typically follows a log-log relationship (Hoffman et al, 2022). Although performance on training objectives—usually next token prediction—does not directly translate to real-world capabilities such as code creation or question-answering, larger models are increasingly used due to the observed association between training performance and practical capabilities (see, e.g., Radford et al., 2019; Brown et al., 2020).

### 2.2.5 Talent

In addition to this triad, talent is also important for the development of frontier AI models. Technical expertise and talent serve as major bottlenecks in the AI industry overall (see, e.g, Gehlhaus et al., 2023). The talent is often concentrated in a few key hotspots of expertise and innovation, as highlighted by the economic literature of technological clusters (see, e.g., Kerr & Robert-Nicout, 2020).

## 2.3 Steps of the supply chain

The AI market is characterized by global reach, complexity, concentration, high fixed costs, and significant investments in research and development (R&D). Roughly from down to upstream, the key steps that make up this supply chain are i) AI laboratories, ii) cloud providers, iii) chip designers, iv) chip fabricators, and v) lithography companies that build the machines used in the fabrication of AI accelerators. See the Pilz' (2023) visualization of the advanced AI supply chain. As featured in the diagram, there are other relevant steps, such as the Core IP, OSAT (Outsourced Semiconductor Assembly and Test), and supplier of key inputs to lithography companies that we are not going to focus on in this report.

### 2.3.1 AI labs and cloud providers

AI labs design, train and deploy frontier AI models. Four major AI labs that are actively engaged in the pursuit of developing Artificial General Intelligence (AGI) are OpenAI, Google DeepMind, Anthropic and xAI. In addition to them, Microsoft, Meta, Apple, and Mistral develop large foundation models.

Major tech companies like Google, Amazon, and Microsoft offer both consumer and business cloud services. Cloud services are widely used in developing and deploying AI models. These platforms enable developers and businesses to access pre-built AI tools, frameworks, and APIs, allowing them to use AI capabilities without needing extensive infrastructure investment.[3]

The AI industry is significantly influenced by big tech in other ways. For instance, Meta created PyTorch by Meta and Google developed TensorFlow, respectively the first and second most widely used AI frameworks for the development of models.

### 2.3.2 The semiconductor industry

The semiconductor industry involves the design and manufacturing of chips, as well as the machines needed to produce them. The process starts with getting silicon from sand and purifying it using specialized chemical methods. Silicon is a key material that allows us to make transistors, which are small electronic parts that can represent binary code and form the core of any computer system. These chips are made from larger pieces called wafers and are carefully designed to fit as many transistors as possible (see, e.g., Proia, 2019).

The semiconductor industry originated primarily in Silicon Valley during the 1960s. Gordon Moore, founder of Intel, famously predicted that the number of transistors on chips

*Overview of the advanced AI supply chain (Pilz, 2023)*

[3] In the next section we cover the relationship between AI labs and big tech companies

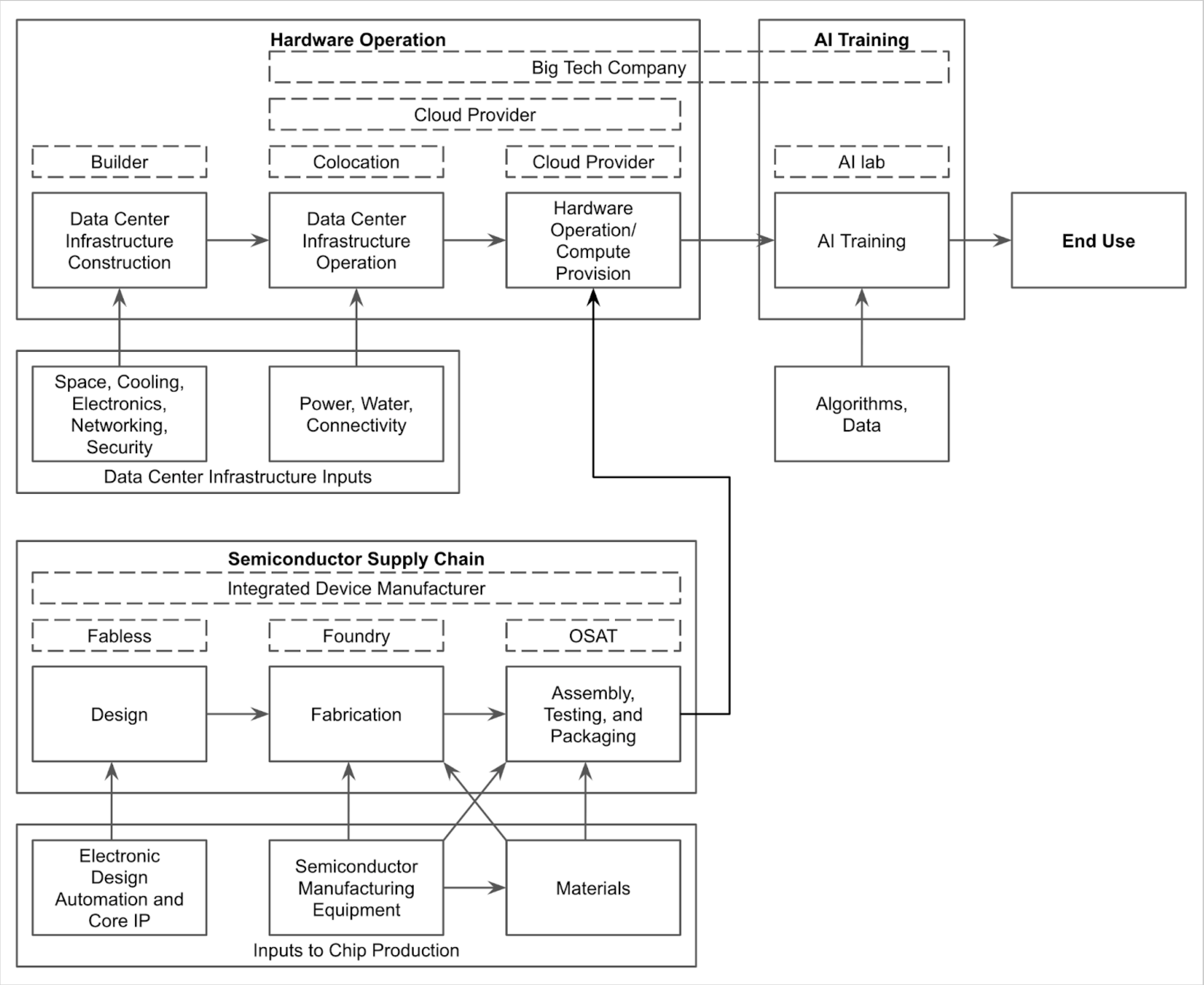

*Overview of the advanced AI supply chain (Pilz, 2023)*

would double every two years, a forecast that became known as Moore's Law. Aiming this, companies were and are focused on reducing the node size of chips, which refers to specific manufacturing processes and the size of the features it can create, usually related to the size of a transistor's gate. These smaller nodes allow for more transistors to be packed closely together, which usually means better performance and efficiency for the chip. While the terminology is not used consistently in the industry, the node size is typically expressed in nanometers and conveys the technological generation of the product.

Early industry giants such as IBM and Texas Instruments were integrated companies, handling everything from machine manufacturing to chip design and fabrication. A significant shift occurred in 1987 with the founding of TSMC (Taiwan Semiconductor Manufacturing Company) by Morris Chang. In collaboration with the Taiwanese government and Philips, TSMC specialized in manufacturing, paving the way for the prevalence of fabless companies, which focus on design while outsourcing fabrication (Chiang, 2023). Nowadays, the manufacturing of semiconductors is mostly concentrated in East Asia, including countries such as Taiwan, Japan, South Korea, and China (see, e.g., Thadani & Allen, 2023).

In the early 2000s, the semiconductor industry saw a shift with the introduction of extreme ultraviolet (EUV) lithography technology, used for the production of advanced chips. While Nikon and Canon were dominant players, the emergence of ASML, a company from the Netherlands, made them a de facto monopoly in the EUV sector, essential for manufacturing the most advanced chips. This underscores how technological advancements can lead to market concentration, influencing the overall industry dynamics.

In this report, the main focus will be on a type of chip called AI accelerator. These chips are specifically designed for tasks related to AI. These chips are optimized for tasks such as matrix multiplications and tensor operations, which are very common in AI training and inference. In contrast, for instance, to CPUs, which prioritize general-purpose processing and are optimized for a broad range of computing tasks including logical, arithmetic, and control operations, AI accelerators are built specially to manage the high-throughput, parallel computations commonly found in machine learning tasks.

NVIDIA's GPUs are the dominant hardware accelerator in the AI industry. Google has its own solution, an ASIC known as Tensor Processing Units (TPUs), available only through cloud services. The semiconductor industry also uses ASICs for particular AI applications such as recommendation systems, computer vision, and natural language processing. Recently, there has been a trend toward creating chips specialized either for training or deployment tasks (Reuther, 2022).

### 2.3.3 Other relevant segments

The semiconductor supply chain includes important segments like CORE IP (Intellectual Property) and OSAT (Outsourced Assembly and Test). CORE IP companies create the basic reusable design of chips, known as IP cores. These designs are licensed to chip designers like Nvidia. Key players in this area are ARM, Synopsys, and Cadence, among others (Design Reuse, 2020).

After wafer fabrication by foundries, Outsourced Assembly and Test (OSAT) companies handle the cutting, assembling, packaging, and testing of chips to turn them into finished products for market release. This stage is crucial for the final quality and efficiency of chips and important companies include ASE Technology, Amkor Technology, and Lam Research. Though CORE IP and OSAT are also important segments, this paper will briefly cover these steps, focusing more on other parts of the supply chain.

# 3. Overview of the integration landscape

We cover mergers and acquisitions worth at least USD 100 million or that otherwise proved especially relevant. Additionally, we will introduce strategic partnerships, joint ventures, investments, disinvestments, and initial merger and acquisition talks that did not succeed. In Section 4, we will note relevant antitrust litigations. In Section 5, we preliminarily discuss what may be its drivers.

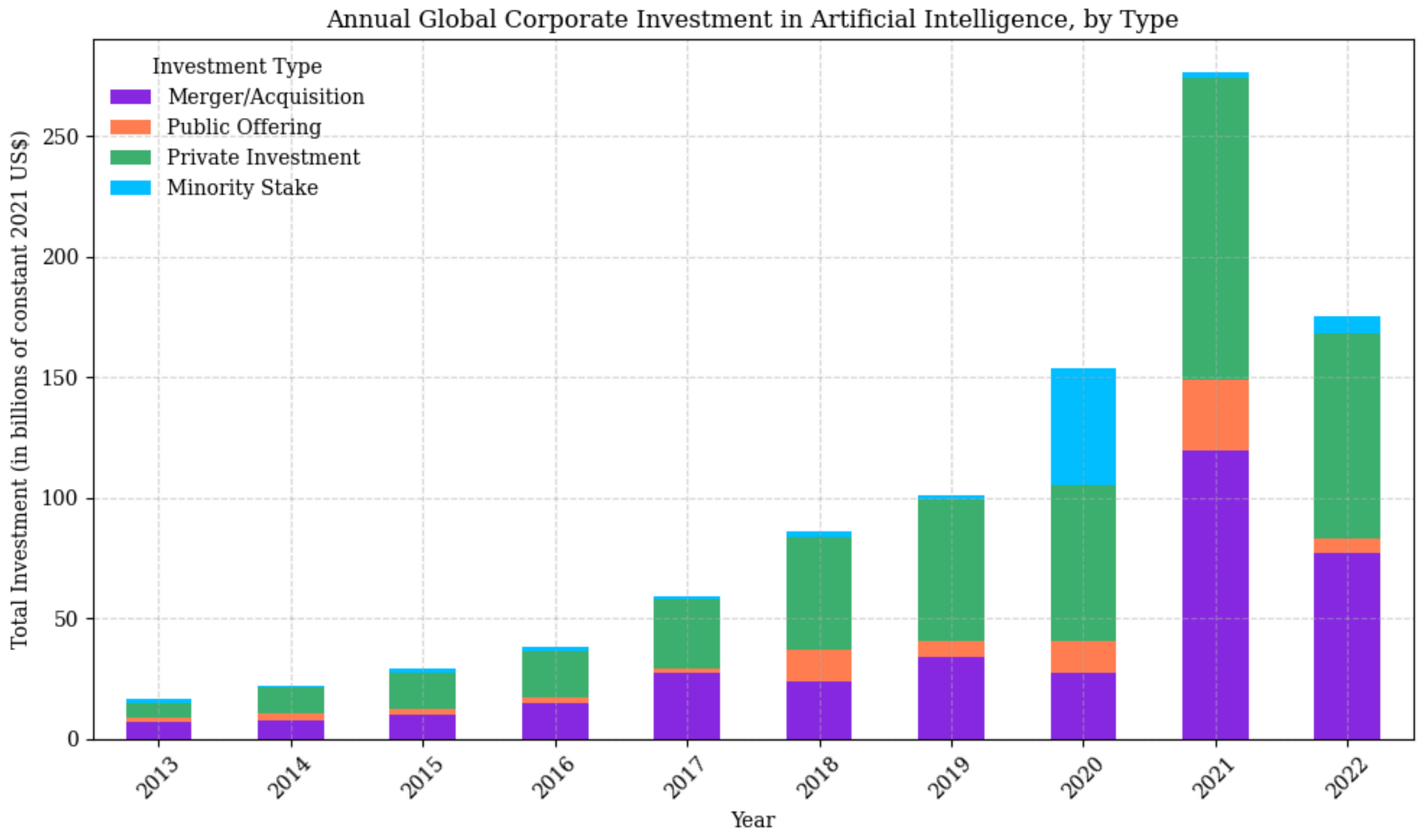

Data source: NetBase Quid via AI Index Report (2023), and is based on Our World in Data (2024)

## 3.1 Working definitions

For this project, we have made preliminary efforts to identify the relevant product markets, trying to be consistent with relevant guidelines in the US and EU. Due to the global nature of these markets, we are not going to focus on discussing geographically relevant markets, though they could prove themselves relevant at times.

In each market, we are going to separate the companies into "frontier" and "non-frontier but relevant". This distinction is useful in two ways: first, it highlights how contestable each of these markets is; second, it may be useful for identifying potential instances of vertically differentiated products.  For tractability, we are not going to cover companies worth less than USD 1 billion dollars as of October 2023 or that are based mainly in China. As noted by, e.g., Jamison (2014) defining relevant markets is a significant, multidisciplinary challenge, especially in dynamic industry with differentiated products and these are provisional definitions.

#### 3.1.1 AI Labs

***Tentative relevant product market definition:*** *Large Language Models (LLMs) with capabilities including but not limited to text generation, text summarization, and code creation.*

In the scope of AI labs, we focused on large language models with varying capabilities, such as text generation, text summarization, and code creation. These models potentially constitute their own relevant product market due to these unique capabilities that are not easily substitutable. We will consider frontier companies in this market as the ones in which the best-performing models are comparable both in capabilities and quantity of FLOPs utilized from GPT-3.5 to GPT-4. Conversely, companies considered relevant but not at the frontier are those whose best models are equivalent in performance to GPT-3 to GPT-3.5.

Frontier models include OpenAI's GPT-4, Anthropic's Claude 3, Meta's LLaMA 3, and Google's Gemini. We also decided to include Inflection's Pi as a frontier. Although its capabilities are more limited by design as it is intended to be a more restricted personal companion,  its underlying model (Inflection-2) is arguably as powerful as the above models (Inflection AI, 2023). This highlights how product differentiation may blur the definition of relevant markets, especially in the case of foundation models that by definition can be used in a wide range of downstream tasks.

Both technical and market reports confirm these categorizations. For instance, in a press release, Inflection AI (2023) recognizes PaLM 540B (that underlies Google's Bard), GPT-3.5 and 4 and LLaMa as its competitors; Mollick's (2023) 'opinionated guide' to use AI in workflows regards GPT-4, Claude and Bard as rough substitutes; and Zhao et al. (2023) point out that GPT-4 and Claude 2 have similar performance as general task solvers in benchmark tests and LLaMA 2 is the best performant open-source model tested. Chiang et al (2023) developed a leaderboard of LLMs, using pairwise crowdsourced comparisons of models, to rank models.

Non-frontier companies include Cohere, which focuses on the integration of LLMs in business processes, and HuggingFace' Hugging Chat, which borrows from several open-source models. It is essential to acknowledge the existence of other classes of models, such as sentence embedding models and multimodal models, as well as technologies that maximize the utility of large language models (LLMs), like Retrieval-Augmented Generation (RAG). Inherently, the fine-tuning of LLMs for specific downstream tasks could constitute a market of its own significance. However, we have chosen not to focus on these other aspects.

### 3.1.2 Cloud providers

***Tentative relevant product market definition:*** *Cloud infrastructure services capable of training LLMs*

We are going to consider frontier cloud providers' infrastructure as the companies with data centers capable of training the frontier foundation models. This includes primarily Google Cloud, Amazon Web Services, and Microsoft Azure. Google's models are trained in-house, Anthropic used Google Cloud but now is transitioning to Amazon and OpenAI's exclusive cloud provider is Microsoft Azure.

We will also consider non-frontier including Oracle. Although there are no reports of what we would consider frontier or non-frontier but relevant LLMs trained in their infrastructure it is reasonable to consider they are contenders for that. For instance, Oracle has set up a partnership with Cohere (Oracle, 2024). Big tech companies such as Apple and Meta also have their data centers capable of training frontier foundation models but do not engage in the B2B cloud market.

### 3.1.3 AI Chip Designers

***Tentative relevant product market definition:*** *Chips used for training and running LLMs*

Regarding chips, our focus was on those used in the data centers of the frontier cloud providers, mainly Nvidia's GPUs and Google's TPUs. As we will further discuss in subsection 3.3.5, these are the only chips that were used by companies to train frontier LLMs. We also considered as non-frontier but relevant: AMD, which is considered as a substitute in, e.g., Shavit, 2023, and EpochAI, 2023; Intel, which has its own AI accelerators (Intel, 2023); and Cerebras chips, which develops AI accelerators with comparable features as of Nvidia and Google's, but arguably has not yet achieved commercial viability.

Additionally, we will consider AI chips developed internally by Meta, Microsoft, Amazon, and Apple as non-frontier but relevant, since they all can reasonably be contenders. We further discuss them in subsection 3.3.4.

### 3.1.4 AI Chip Fabricators

***Tentative relevant product market definition:*** *Companies capable of producing aforementioned chips with sufficient node precision*

For chip fabricators, we looked at the companies that produce AI accelerators that the cloud providers use in their data centers. Nvidia's GPUs are largely produced by TSMC, while the fabricator for Google's TPUs is undisclosed, though Samsung is a probable source (TrendForce, 2023). Both TSMC and Samsung are producing 3nm chips and plan to produce 2nm until 2025. We will also consider Intel and GlobalFoundries as non-frontier but relevant.

### 3.1.5 Lithography companies

***Tentative relevant market definition:*** *Companies capable of producing machines used by chip fabricators with sufficient node precision*

The semiconductor lithography equipment market as a whole has an estimated annual revenue of USD 24.66 billion dollars (Mordor Intelligence). ASML provides Extreme Ultraviolet (EUV) and Deep Ultraviolet (DUV) lithography machines for leading chip manufacturers such as TSMC, Samsung, and Intel. With its offerings of advanced lithography systems, ASML has established itself as the de facto monopoly in this sector. On the other hand, Japanese firms Nikon and Canon are recognized as potential competitors within this market. Although they haven't mastered EUV technology, both companies offer DUV systems suitable for fabricating chips that do not require high-node precision

### 3.1.5 Additional notes

We will also follow the following taxonomy to classify different types of integration between companies:

| **Type of Integration** | | **Description** |
|---|---|---|
| **Vertical** | | A company integrates with another operating within the same production process but at different stages. |
| | Forward | Moving closer to the end customer (e.g., a manufacturer opening retail outlets). |
| | Backward | Moving closer to raw materials (e.g., a car manufacturer acquiring a tire company). |
| **Horizontal** | | Increased participation in the same industry. |
| **Conglomerate** | | Increased participant participation in unrelated activities |
| | Pure Conglomerate | New participation in industries that have nothing in common (e.g., a food company merging with a shoe company). |
| | Mixed Conglomerate | New participation in product or market that is perceived by consumers as complementary (e.g.: a shoes company merging with a socks company) |

Each kind of integration can be also classified as: i) integration by acquisition (one company buys another; ii) integration by merge (i.e., two companies create a new one); iii) integration by expansion (i.e., a company starts participating in new industries without acquiring/merging another company); iv) quasi-integration (characterized by minority stake and/or partnership with exclusivity clauses on key inputs). We tried to be generally consistent with the nomenclatures adopted by industrial organization textbooks such as Tirole (1987) and Shy (1996).

It is important to note sometimes it can be challenging to define which kind of integration a merger or acquisitions is, since it depends on the definition of the relevant market.

## 3.2 Integration in the AI supply chain

The AI supply chain, especially if we only consider the frontier AI supply chain, is highly horizontally concentrated. There is only one frontier lithography company (ASML), only one or two frontier AI accelerator fabricators (TSMC and Samsung), and only two designers of cutting-edge AI accelerators (Google and Nvidia), with only one supplying the broader market. There are perhaps three to fifteen frontier AI labs, such as Google Deepmind, Meta, Anthropic, Inflection, and OpenAI. However, there is significant horizontal shareholding. Besides owning Deepmind, Google also invested in Anthropic. Microsoft has a strategic partnership with OpenAI. Microsoft also invested in Inflection and hired a substantial number of its founding members to work at the newly created Microsoft AI.

Regarding vertical integration, from approximately the 1950s to the 2000s, there seemed to be a general trend toward less vertical integration in the semiconductor industry, notable through fabless companies becoming more common. However, with the challenges of maintaining pace with Moore's Law upstream and the rise of big tech companies downstream, there seems to be a swing back towards increased vertical integration.

In the industry, there is a prevalence of strategic partnerships that often involve exclusive clauses or minority stakes. These arrangements closely resemble quasi-integration, particularly concerning the provision of key inputs or technology licensing. From the 25 mapped companies, we considered that nine companies are present in more than one of the five relevant markets mapped.

| | **Lithography Companies** | **AI Chip Fabricators** | **GPU Designers**[4] | **B2B Cloud**[5] |
|---|---|---|---|---|
| All companies | | | | |
| Number of companies (>1% of the market share) | 3 | 5 | 13 | 8 |
| Top 1 Concentration Ratio | 60% | 56% | 90% | 34% |
| Top 3 Concentration Ratio | 100% | 78% | 100% | 66% |
| Rough estimate of HHI | 4600 | 3800 | 8150 | 1750 to 1936 |

### 3.2.1 Lithography Companies

ASML has significantly expanded its market share by being the first and so far the only company to develop EUV (Extreme Ultraviolet) technology. There have been no mergers and acquisitions in the industry that significantly impacted the horizontal integration of the lithography industry in at least the last 25 years.

The Japanese companies Nikon and Canon, once significant players in the lithography market, have fallen behind in technological innovation. Canon lagged by two generations, failing to master immersion lithography, while Nikon remained a generation behind, still selling immersion DUV machines but abandoning attempts at EUV development.

ASML has acquired and invested in numerous of its suppliers of key inputs of lithography machines. As part of its strategy to develop Extreme Ultraviolet (EUV) technology (ASML, 2012), ASML purchased Cymer, a supplier of light sources used in lithography processes. In 2016, ASML bought Hermes Microvision, a Taiwanese company that develops electron beam inspection technology used to identify defects in advanced integrated circuits (ASML, 2023).

ASML and ZEISS have a long-standing partnership, specially through the ZEISS subsidiary focused on technologies used in the semiconductor industry, Carl Zeiss SMT. In

[4] High-end GPUs, which excludes various other forms of AI accelerators that should be potentially included in the same relevant market.

[5] Synergy Research Group (2022). Lower (upper) estimate calculated assuming that the categories "next 20 largest" and "others" consider the lowest (highest) potential concentration.

2016, ASML bought 24,9% of Carl Zeiss SMT for EUR 1 billion (ASML, 2023). While ASML focuses on the lithography machines, Zeiss specializes in the optics that go into these machines. As stated in the original announcement from the companies, “the main objective of this agreement is to facilitate the development of the future generation of Extreme Ultraviolet (EUV) lithography systems due in the first few years of the next decade” and it has been successful.

Canon has recently introduced a nanoimprint lithography (NIL) machine, boasting capabilities for 5nm chip production and tracking 2nm production eventually. The market acceptance of this new offering from Canon remains an open question as it awaits commercial adoption. Nikon, ASML and Carl Zeiss ran intellectual property litigation against each other for years and it ended up with a legally binding cross-licensing agreement (ASML, 2019).

In summary, the market consolidation in the lithography market seems to be mainly driven by the natural growth of ASML given their technical advantage over the competitors and acquisitions of and alliances with key suppliers and customers.

### 3.2.2 AI Chip Fabricators

There were no major horizontal integrations involving TSMC, Samsung, Intel, or GlobalFoundries that occurred in the industry over the past 25 years.

As TSMC's market share in sub-7nm nodes reached nearly 90% and reportedly had a general market share of 58% as of the last quarter of 2022 (The Motley Fool, 2022), they are widely regarded as the most advanced foundry company. In 2022, they started fabricating 3nm process nodes commercially and currently are developing technology for 2nm processes (TSMC, 2023).

In 2022, Samsung was also able to produce a 3nm node process and is currently starting to sell them commercially, reportedly seducing a large contract for producing specialized data center chips (Pulse, 2023). The South Korean conglomerate is also trying to keep pace with TSMC in the development of the 2nn nodes process. Intel has been losing market share in the world market. Noteworthily, they have lost significant contracts with Apple and AMD. It is worth noting there are notable licensing agreements, such as those concerning 12nm foundry technology from Samsung to GlobalFoundries (WikiChip, 2018)

In 2012, ASML set up a consumer co-investment program with the goal of developing EUV technology that could be employed at scale for the fabrication of chips. Its three larger customers at the time bought a total stake of 23% of ASML. The program was enabled through a synthetic share buy-back. We put more details of this partnership, which was fundamental to the development of  EUV technology on which today's frontier AI accelerators depend, as a case study in the appendix.

### 3.2.3 AI Chip Designers

Nvidia has a significant market dominance in the Graphics Processing Unit (GPU) market, with more than 80% of this market (Jon Peddi Research, 2023). Their CUDA environment, which helps the optimization of machine learning training and inference for their specific hardware, is considered by market analysts as a significant advantage. In 2020, Nvidia acquired Mellanox for $7 billion. As Nvidia is a leading developer of chips used in data centers, and Mellanox specializes in complementary data center technologies, this acquisition is an example of mixed conglomerate integration.

AMD is one of the top contenders of Nvidia. They have been investing in open source infrastructure, being a partner of the Torch Foundation, which develops the PyTorch framework, and recently buying the company Nod.ai.

Another significant player in this market is Cerebras, which produced the first 1-trillion-transistors chips. Big tech companies have also been engaged in the designing of AI chips, typically for their in-house operations, as we will convert in the vertical integration subsection.

Generally, companies that both fabricate and design chips are called  Integrated Device Manufacturers. They usually fabricate their self-designed chips as well as sell them to other (fabless) companies. Intel is a major US integrated manufacturer and has a division specializing in foundry services for other companies. Some of its major clients for these services are Meta, Amazon, Cisco, and MediaTek.  TSMC and Nvidia have established a strong partnership and are the dominant players in chip manufacturing and design, respectively.  In March 2023, Nvidia announced CuLitho, a software to improve computation lithography. The company has partnered up with ASML, TSMC, and Synopsis “to accelerate the design and manufacturing of next-generation chips" through integrating computation processes of lithography in GPUs (NVIDIA, 2023). One of the stated goals is to push towards making chips with nodes of 2nm and less. This partnership does not include a company acquiring participation in another, however.

### 3.2.4 Cloud providers

The cloud market at scale is dominated by AWS, Microsoft Azure, and Google Cloud, and, respectively with 34%, 21%, and 11% of the market. To the best of our knowledge, there have been no mergers and acquisitions in the industry that significantly impacted the horizontal integration of the cloud industry in at least the last 25 years. There have been, however, noteworthy conglomerate acquisitions, especially focused on cloud providers expanding their portfolio of services.

For instance, in 2019 Google bought Looker, a big data analytics platform, for 2.6 billion dollars to expand Google Cloud's offerings in the business intelligence segment. Also in 2019, IBM bought Red Hat, a software company focused on open-source enterprise software, for 34 billion dollars. In 2018, Microsoft bought GitHub, which provides a platform for software development, for 7.5 billion dollars, integrating it into the Azure cloud platform. Later on, GitHub data was used to train foundation models.

Cloud providers are actively also involved in chip design, enabling them to have more control over their hardware and optimize it for specific applications. Google's Tensor Processing Units are available for use through the Google Cloud Platform. Similarly, Amazon develops its Trainium and Inferecium chips, available through Amazon Web Services. Reportedly, Microsoft has been internally developing AI accelerators. Microsoft CTO confirmed that they are investing in semiconductor solutions, but did not provide further details.

Google's Tensor Processing Unit (TPU) is specifically designed to enhance AI computations and model inference and is one of the most used AI accelerators (Reuther et al, 2021). TPUs are not physically available in the market, one can only use them through Google Cloud (Google, 2023).

Amazon designs chips that are specialized for cloud operations, focusing on the specific requirements of its cloud services and infrastructure (Amazon, 2023).

Microsoft's acquisition of a chip-design startup in early 2023 indicates their interest in developing AI chips (Fool, 2023). While specific details are not available, reports suggest their focus on AI-related chip development (CoinTelegraph, 2023).

It is also worth noting that, although they are not cloud providers, Meta also announced in early 2023 their AI chip project, Meta Training and Inference Accelerator (MTIA), designed to improve efficiency for recommendation models used in serving ads and content on news feeds (SourceAbility, 2023). Apple has transitioned from being a customer of Intel to designing its own chips for its latest laptop to having greater control over the performance and integration of its hardware with its software ecosystem, but there is no information on the design chips specialized in AI tasks.

### 3.2.5 AI Labs

The only significant merger and acquisition that increased horizontal integration of frontier AI labs that we are aware of is Google's acquisition of DeepMind in 2014. However, as they shared different focuses in AI technology, Google Brain and DeepMind remained independent until 2023 and the key aspect of this deal was Google providing compute at scale to DeepMind, we will treat it as vertical integration.

AI labs have a large demand for compute power to train large foundation models. There is a need to have many frontier AI computers closely connected in a data center, as is explored in the Pilz and Heim (2023) report.

As discussed above, there are only three major companies currently capable of supplying at scale advanced AI accelerators to their customers: Google (Google Cloud), Microsoft (Azure), and Amazon (AWS). Besides their use for internal AI development, these three companies have set strategic partnerships with top AI startups. Especially as there is a constrained supply of chips, the allocation of current capabilities is increasingly important and set by these strategic partnerships.

#### 3.2.5.1 Google Cloud ↔ DeepMind

Deepmind was founded in 2010 and was acquired by Google in 2014. The exact purchase price was not disclosed, but it was reported to be between USD 500 million and USD 650 million (Efrati, 2014; Gibbs, 2014; Chowdhry, 2015). Facebook was also interested in purchasing at the time. Reportedly it is unclear why the conversations with Facebook didn't advance (Efrati, 2014). DeepMind's acquisition by Google was reportedly led by Google CEO Larry Page. One of the conditions for the purchase was the creation of an ethics board, as DeepMind was created with strong AI safety concerns. It is uncertain if DeepMind already relied on Google Cloud before the purchase. Since the acquisition, DeepMind has been leveraging Google Cloud's infrastructure and services for its AI research and development.

DeepMind became a wholly owned subsidiary of Google's parent company Alphabet Inc. after Google's corporate restructuring in 2015. In April 2023, Google Brain and DeepMind merged to form a new unit named Google DeepMind with the goal of accelerating the development of general AI. Reportedly this was an answer to OpenAI's breakthrough with ChatGPT (VentureBet, 2023).

#### 3.2.5.2 Microsoft's Azure ↔ OpenAI

OpenAI, established as a capped-profit company subsidiary of a non-profit organization, has received support from Microsoft since 2019 (Open AI, 2019). In 2023, this commitment from Microsoft with OpenAI was renewed with an investment of 10 billion dollars (Open AI, 2023). Microsoft Azure, which serves as the sole cloud provider for OpenAI. The AI lab does not maintain any data centers of its own. In addition, Microsoft possesses exclusive access to the parameters of the GPT-3 and GPT-4 models, incorporating them into a diverse range of its products (CNBC, 2023).

#### 3.2.5.3 Microsoft's Azure ↔ Inflection

Inflection was co-founded by Mustafa Suleyman of Google DeepMind and Reid Hoffman of LinkedIn focusing on creating consumer-facing AI products like the chatbot Pi. . As an early investor, Microsoft provided Inflection with access to Azure infrastructure and built a cluster of around 22,000 H100s. The founders later established Microsoft AI, a division that consolidates consumer AI initiatives from Bing and Edge, enhancing Microsoft's consumer AI offerings.

Now, Inflection concentrates on growing "AI studio", developing custom AI models for commercial clients and making them available through cloud services such as Azure (Reuters, 2023; Inflection, 2024).

#### 3.2.5.4 Google Cloud ↔ Anthropic

In February 2023, Anthropic partnered with Google Cloud as its cloud provider (Anthropic, 2023). Dario Amodei, Anthropic CEO, has said that "We've been impressed with Google Cloud's open and flexible infrastructure." (Edgeir, 2023) Anthropic will be able to use GPU and TPU available in Google's clusters and Google has invested USD 300 million in Anthropic (Financial Times, 2023).

#### 3.2.5.5 AWS ↔ Anthropic

Amazon and Anthropic have established a substantial partnership. Since 2021, Anthropic has been a client of Amazon (AWS, 2023) Amazon, through AWS, has facilitated access to Anthropic's generative AI model Claude for AWS customers via Amazon Bedrock. On September 25, 2023, Amazon announced an investment of up to $4 billion in Anthropic to bolster the development of language models like Claude 2 using AWS and its specialized chips (Amazon, 2023). According to the announcement, this investment is part of a broader strategic collaboration aimed at advancing safer generative AI technologies and making Anthropic's future foundation models widely accessible through AWS.

#### 3.2.5.6 AWS ↔ HuggingFace

The two companies have set a strategic partnership and now Amazon offers models available in the HuggingFace (HuggingFace, 2023). This partnership is specially focused on training and deployment of AI models in the HuggingFace platform using Amazon Web Services cloud computing services In the announcement, they focused on the benefits customers may get from this partnership. They don't mention directly that AWS will provide HuggingFace compute for training their own models (e.g.: HuggingChat).

#### 3.2.5.5 Cohere ↔ Oracle

In June 2023, Cohere announced a partnership with Oracle to enhance AI services in the companies' platforms. They announced that they are working together to ease the training of specialized large language models for enterprise customers while ensuring data privacy during the training process. Cohere's generative models are integrated into Oracle Cloud Infrastructure (Oracle, 2023).

### 3.2.6 AI Chip Designers ↔ AI Labs

Besides Alphabet and Microsoft (which we discussed above considering them mainly as cloud providers), there has been limited integration of AI chip designers and AI labs. In 2018, NVIDIA created its Toronto AI Lab. Their focus "lies at the intersection of computer vision, machine learning, and computer graphics." (NVIDIA, 2023). NVIDIA has not however acquired companies focussed on AI model development.

Meta has been developing chips specifically for inference in AI tasks such as computer vision and recommendation systems. Other companies that have significant involvement in AI model research and chip designing are Samsung and Apple, though they are not frontier players in any of these industries.

Samsung has seven research centers dedicated to AI across countries as South Korea, the United States and Russia (Samsung, 2023)

Apple has a significant investment in natural language processing for its voice system, Siri, and has been reportedly trying to integrate foundation models developed internally into the operational systems of its products. The extent to which Apple is strategically investing in this is not clear and the company is famous for being one of the most secretive in Silicon Valley. They reportedly have been in talks with both Google and OpenAI to integrate their models on iPhone (Euronews, 2023).

### 3.2.6 Other

In 2016, Softbank, a Japanese holding with various investments, ranging from telecom services to internet-based businesses, acquired the semiconductor intellectual property (IP) company ARM for approximately $31 billion—a conglomerate integration.

From 2020 to 2022, NVIDIA tried to acquire ARM, the leading licenser of CPU designs. The acquisition could have allowed NVIDIA to integrate ARM's CPU technology with its AI accelerators, but the deal faced antitrust challenges and NVIDIA ultimately gave up on acquiring ARM.

# 4. Antitrust in the AI supply chain

There has been growing interest in more active antitrust measures in the technological sector. In this section, we highlight how this has been impacting relevant steps of the AI supply chain. We generally observe concerns of abuse of dominance, such as bundled sales and exclusive dealing, in the semiconductor industry and big tech companies. There have been some noteworthy merge controls in the semiconductor industry that we will describe, but none in the cloud providers or AI products directly.

## 4.1 Lithography and semiconductors

Until now, there has been limited antitrust litigation in the companies upstream of the AI supply chain. No acquisition by ASML and TSMC, respectively the most advanced lithography company and the most advanced foundry company in the world, has been challenged by US or EU authorities, and they were not adversely affected by any other kind of antitrust litigation in the past 25 years.

In the upstream part of the industry, the most noteworthy antitrust case was from companies that supplied non-lithography semiconductor manufacturing equipment to chip fabricators. From 2013 to 2015, Applied Materials and Tokyo Electron attempted to merge for USD 29 billion, which raised concerns from authorities as this would be a horizontal integration of, respectively, the first and second-largest company in this market segment. The merger was eventually abandoned by the companies after the Department of Justice of the US raised competition concerns and rejected their proposed remedies (Department of Justice, 2015).

The chip designing industry is receiving increasing attention from antitrust authorities. The FTC challenged Nvidia's acquisition of Arm Limited by USD 40 billion because of concerns that NVIDIA would gain excessive market power as it would have incentives to foreclose the licensing of Core IP owned by Arm to other chip designers. In 2022, NVIDIA terminated the proposed acquisition of Arm (FTC, 2022; TechCrunch, 2022). This acquisition was also under scrutiny in the European Union under similar concerts (Reuters, 2021). Additionally, the European Union is reported to have launched an early investigation into suspected anti-competitive abuses by Nvidia in the AI chip market (Tech Going, 2023) and, in France, Nvidia's offices were raided by the country's antitrust authority over suspicions of anticompetitive practices (Forbes, 2023). In the US in 2008, Nvidia settled a GPU antitrust class action lawsuit in 2008, which alleged a price-fixing conspiracy with ATI to fix, raise, maintain, and stabilize prices of graphics processing chips and cards (Bit-Tech, 2008).

In 2005, Advanced Micro Devices (AMD) opened a private antitrust lawsuit against Intel around allegations of anticompetitive practices in the x86 microprocessor market. Filed in June 2005 in the United States by AMD, the lawsuit accused Intel of engaging in illegal practices to maintain a monopoly over the market, including offering rebates to companies for purchasing most of their microprocessors from Intel, and retaliatory actions against customers who engaged with AMD. The case culminated in a settlement in 2009, where Intel agreed to pay AMD $1.25 billion and adhere to a set of business practice provisions to enhance competition in the microprocessor market (AMD, 2009). Afterward, Intel also reached a settlement agreement with the FTC, which prohibited the company "from using threats, bundled prices, or other offers to exclude or hamper competition or otherwise unreasonably inhibit the sale of competitive CPUs or GPUs" (FTC, 2010).

The legal battle was part of a broader global scrutiny of Intel's practices. South Korea and the European Union also investigating Intel's market behavior. In Japan, the Fair Trade Commission (JFTC, 2005) took action against Intel in 2005, accusing the company of offering rebates to five prominent PC makers—Fujitsu, Hitachi, NEC, Sony, and Toshiba—on the condition that they limit or cease purchases from Intel's competitors, primarily AMD (CNET, 2005). Following this, Intel agreed to a cease and desist order (NetworkWorld). Around the same time, South Korea's antitrust authority, the Korean Fair Trade Commission (KFTC), initiated an investigation into Intel's practices in 2005, culminating in a fine of Won 26bn ($25m) in 2008 for abusing its dominant market position in the country (Computer World, 2018). The European Commission, too, was probing Intel's market behavior, and in collaboration with Japanese authorities, was investigating possible antitrust violations.

Another major block to merge in the semiconductor industry has been the block to the proposed merger of Broadcom and Qualcomm. The proposed $117 billion merger between Singapore-based Broadcom Ltd and U.S.-based Qualcomm Inc faced severe scrutiny from U.S. authorities, leading to its blockage by President Trump due to national security concerns, particularly fearing an erosion of U.S. mobile technology leadership to China's advantage (Reuters, 2018). Despite Broadcom's attempts to alleviate concerns by pledging to redomicile to the U.S. and not sell critical national security assets to foreign entities, the merger was halted, reflecting U.S. efforts to safeguard national and technological security in the semiconductor industry (PCMag, 2018).

There has also been a major price-fixing scandal in the semiconductor industry. The DRAM cartel scandal emerged in the early 2000s with multiple major manufacturers of dynamic random-access memory (DRAM) being implicated. The US Department of Justice initiated a probe in 2002, responding to claims from US computer makers like Dell and Gateway regarding inflated DRAM pricing impacting their profits (Department of Justice, 2005). Samsung, Hynix, Infineon, Micron Technology, and Elpida pleaded guilty to their involvement in a cartel spanning 1998 to 2002. On a global scale, European antitrust regulators fined nine semiconductor manufacturers over €331 million in 2010, reflecting actions that took place in 2002 (European Commision, 2010). The scandal saw criminal fines totaling more than $730 million against the DRAM cartel members, marking at the time the second-largest total amount of fines ever imposed in a U.S. criminal antitrust investigation (EDN, 2007).

## 4.2 Cloud and AI

While big tech companies have been in increased scrutiny by antitrust authorities, especially for anti-competitive behavior in price-setting and damaging its competitors in their platforms, there has been no noteworthy antitrust litigation directed impacting the development of frontier AI systems. DeepMind acquisition by Google, for instance, was approved without conditions, with no significant information publicly available of antitrust authorities raising concerns about this deal.

## 4.3 Policy: sanctions, tensions, and subsidies

From their inception, the semiconductor and AI sectors have often been viewed as strategic markets for government involvement. For example, Silicon Valley's growth was partly fostered by DARPA contracts (see, e.g., Heinrich, 2002). Recently, both the U.S. and EU governments have shown a renewed focus on subsidizing the chip industry. We overview these actions for completeness, but they are not the focus of this paper.

### 4.3.1 Sanctions and geopolitical tension

The semiconductor industry has faced significant tension. In 2019, due to national security concerns, the US imposed sanctions on China's Huawei and pressured other countries not to adopt their 5G technology. In 2023, the US imposed further restrictions on the Chinese technology industry, banning the export of advanced AI accelerators to China and blocking the use of US technologies. The US also pressured allied countries crucial to the semiconductor supply chain.

In response, China launched initiatives to expand its autonomy in the semiconductor industry. The key open question is how long it will take China to develop EUV technology or other comparable technologies independently.

### 4.3.2 Subsidies and industrial policy

Both the US and the EU passed Chip Acts, multi-billion dollar subsidy plans to foster the development of the semiconductor industry within their jurisdictions. The US Chip Act is a $280 billion plan to boost semiconductor research and production in the US. It offers $52 billion for chip manufacturing, $24 billion in tax credits for chip tools, and $200 billion for scientific research and innovation. The act also aims to enhance US national security.(Zimmerman, 2022)

The EU Chip Act is a €43 billion ($47 billion) project to support Europe's semiconductor technology and uses. It aims to enhance chip innovation, understand global supply chains, and fill the talent gap in the field, while also targeting Europe's digital and environmental goals (European Commission, 2023).

The Netherlands has ASML, the top maker of chip-making machines. For national security reasons, the Dutch government has limited its sales to some countries, including China (Associated Press, 2023).

Japan's Rapidus, formed in 2022 with the backing of eight major Japanese firms (Denso, Kioxia, MUFG Bank, NEC, NTT, SoftBank, Sony, and Toyota), aims to make advanced 2-nanometer chips by 2027. Rapidus has a tech deal with IBM and received an extra $2 billion from the Japanese government (Kizuna, 2024).

South Korea introduced the "K-Semiconductor Strategy" with a $280 billion investment to boost its semiconductor sector through R&D, subsidies, and tax benefits until 2030. It focuses on national security and enhancing 5G supply chains. The "Semiconductor Cluster" project centers around SK Hynix and Samsung campuses to advance chip technology, backed by government support (Kim et al., 2023).

Taiwan leads in chip production, mainly due to TSMC, the top chip-making company. TSMC makes chips for major tech brands and has invested heavily in advanced chip processes. The Taiwanese government actively helps semiconductor companies secure land, water, and electricity (Chiang, 2023). However, they face political issues from China, which views Taiwan as its own.

China's strategy for semiconductor dominance includes the "China 2025 Plan," targeting 70% self-reliance in semiconductors by 2025. The "Big Fund," initiated in 2014, offers $21 billion in state-supported funds to finance domestic chip endeavors and encourage the acquisition

of foreign expertise and technology. Additionally, a forthcoming $143 billion investment package aims to further enhance China's semiconductor capabilities, prioritizing the production and innovation of advanced chips (Allen, 2023).

# 5. Potential drivers

Several potential reasons may make companies in the AI supply chain vertically integrate, create strategic partnerships, or refrain them from doing that. In this section, we provide an overview of what may be driving these patterns.

## 5.1 Synergies

Vertical integration and partnerships can lead to cost savings and operational efficiencies by consolidating various stages of the supply chain. This integration potentially allows companies to leverage shared resources, infrastructure, and expertise. As the AI industry has high fixed costs, this is probably one of the main drivers.

Moreover, vertical integration helps in reducing transaction costs and improving coordination between different stages of production. This is crucial to avoid problems in R&D projects, as mentioned by Acemoglu et al. (2010). In the AI supply chain, this is seen in the interaction between companies that own manufacturing facilities and chip designers. These transaction costs are both contractual and technological. For instance, companies need to communicate and coordinate closely to develop e.g. a new chip design that utilizes EUV technology to the fullest. An example of that is the partnership between ASML, TSMC, NVIDIA, and Synopsis in developing cuLitho, a software tool being developed to use NVIDIA's GPUs to optimize ASML's lithography technology.

Furthermore, vertical integration and partnerships may also allow companies to acquire talent and technology capabilities by integrating complementary expertise from different stages of the supply chain. This may be an explanation for large technological conglomerates that overreaches a wide range of industries (Chen, Elliot & Koh, 2023) and may also play a role in their recent partnerships and acquisitions of AI labs.

Vertical integration serves as a strategic move for securing essential inputs necessary for AI development. Given the inelastic short-term supply of frontier AI accelerators, companies strive to ensure access to these essential inputs. By taking direct control over various steps of the supply chain, companies can mitigate the risks associated with supply chain disruptions, as highlighted in Elliot & Golub (2022). Many companies seek assurance that they will have access to key inputs, such as compute, because they believe the supply is constrained in the medium run. To secure these resources, they either gain more control over the supply chain or establish strategic partnerships with key suppliers. This dynamic may intensify due to the perceived winner-takes-all nature of the AI industry.

However, as a company grows bigger, the benefits from increasing its size decrease since it becomes harder to manage. Even though this can be lessened by having different corporate structures where subsidiaries operate independently, it remains a major factor preventing companies from integrating too much, both in the AI industry and across the economy. As companies are not capable of specializing in everything but still seek to accumulate capabilities, they potentially see quasi-vertical integration as a flexible solution for that. This potentially is important to understand the strategy of big tech companies such as Alphabet and Microsoft in the AI industry: they seek to have a stake in major AI labs and integrate their technologies in their diversified portfolios of products, but leaving substantial autonomy to the emergent companies.

## 5.2 Strategically harden competition

Firms may aim to strategically create entry barriers in a market to maintain dominant positions. By doing so, they deter potential competitors from entering the market, thus preserving their market share and potentially their pricing power.

In a similar vein, companies may attempt to foreclose access to essential inputs for other firms, as discussed by Patrick & Tirole (2007). The Federal Trade Commission has blocked the acquisition of Arm Limited by Nvidia mainly under this concern (FTC, 2021).

Furthermore, by retaining control over key areas of the business, firms can avoid sharing sensitive data that may otherwise be necessary in more collaborative arrangements. This helps maintain a competitive edge and safeguard proprietary or sensitive information from potential rivals, especially in high-tech sectors, as discussed by Barrera (2019). Firms may also engage in killer acquisitions and capability hoarding to further secure their competitive positioning by either acquiring potential competitors or hoarding critical capabilities to prevent others from accessing them (Cunningham et al, 2019; Boa et al, 2023).

## 5.3 Governmental action or industry reaction

Governments may offer incentives to encourage vertical integration, especially in strategic industries. These incentives can range from tax benefits and subsidies to preferential treatment in procurement or regulatory advantages. This may be an increasingly important driver as the semiconductor industry is increasingly seen as a matter of national security.

Integration may be avoided in the AI industry to mitigate antitrust concerns. Companies involved in the industry may be wary of controlling too much of the supply chain because it could potentially be a concern raised by antitrust authorities. The importance of this is directly impacted by their expectations that this will be an issue for regulators, as discussed by O'Keefe (2021).

Compliance with specific regulations, such as data privacy or security requirements, can be facilitated through vertical integration. The integration allows companies to have better control over data flows, risk management, and adherence to regulatory frameworks such as the EU General Data Protection Regulation (Gal & Aviv, 2020; Carugati, 2023), which may have a detrimental effect on smaller firms and competition overall (Campbell et al, 2015)

## 5.4 Other reasons

As the AI industry is still in a nascent stage, the market for specific services is not that well developed and it is difficult to do major, impersonal transactions since there are no established ways of working. This scenario often drives companies towards vertical integration to secure and streamline operations. Some authors have argued that in early-stage development of general-purpose technologies it is common to see a high degree of vertical integration because of this, followed by vertical separation as markets develop.

Additionally, past business decisions, investments, and established relationships significantly influence the inclination to pursue vertical integration. Companies may choose this path to build upon existing capabilities, intellectual property, or market positioning, thereby leveraging established foundations for growth or competitive advantage. Through this lens, both the early stage of the industry and historical path dependencies play crucial roles in shaping the strategic choice towards vertical integration in the AI sector.

Finally, the patterns we see in the industry may be driven by a growing belief, especially in AGI labs, that the frontier AI industry will exhibit a winner-takes-all dynamic. For instance, Dario Amodei, CEO of Anthropic, mentioned that the current cost of training leading AI models is around $100 million but is expected to escalate to as much as $5 billion to $10 billion by 2025 or 2026. He attributes this sharp increase to the scaling laws of AI (Decoder, 2024). The continuous escalation of compute clusters will only raise the fixed costs necessary to enter the industry.

# 6. Closing remarks and open questions

We have the challenge of understanding the dynamics of the quickly changing AI industry. As with the rest of the hardware industry and software industries (e.g., Tirole, 2023), there are significant barriers to entry associated with relatively low marginal costs in multiple steps of the AI supply chain which may give companies substantial market power. Additionally, we may observe network effects in the industry as well as interoperability issues. The relevance of each of these aspects for each of the steps of the AI supply remains an open question to be sure to what extent the economic theoretical framework needs to be readapted — or reinvented — to fully understand the dynamics of these new industries, similarly to what happened with the rise of the digital economy.

It is clear, however, that the effectiveness of AI regulatory proposals is likely to be heavily impacted by the structure of the AI industry. This final section aims to lay down open questions that can be useful to best understand these implications, paying special attention to vertical relationships. This complements already proposed research agendas such as Siegmann (2023) and Chapter 4 of Winter et al (2021), which respectively pose questions for economists and lawyers regarding AI governance and safety.

Many of the regulatory implications of having more vertical integration in these industries are related to possible trade-offs between competition and safety that can appear in different contexts in the industry. For instance, if we accept the argument that we should decelerate AI development to allow time to understand and address its associated risks, then employing antitrust policies to increase industry competition might be counterproductive. As this would typically fall outside the mandate of antitrust authorities, this may suggest a demand for structural remedies by regulators, besides behavior remedies.

In this sense, there may be tension in looking to enhance short-term consumer welfare and economic efficiency with the mitigation of risks that may arise from frontier AI systems. This trade-off may be less salient if we think about long-term effects on the well-being of consumers.

There are, additionally, national security concerns that may play a role in this. Foster & Arnold (2020) already elaborated how there may be tension between breaking up big tech companies because of their market power and national security. As O'Keefe (2020) pointed out, this attention between focusing on national security and economic efficiency has already been a central part of significant litigations such as AT&T. These concerns also played a role in the block of the merger of Qualcomm-Broadcom.

Noticeably, the regulatory implications of having more horizontal integration can be substantially different from that of having more vertical integration. Hua & Belfield (2020) and O'Keeffe (2021) already explored these horizontal antitrust considerations. However, considerations for vertical integration can significantly differ. Antitrust authorities typically exhibit more leniency toward vertical integration than horizontal, recognizing its potential to boost welfare through efficiency gains. Vertically integrated companies often benefit from shared capabilities across similar economic activities, gain efficiencies from economies of scale, and arguably tend to invest more in research and development as well as in safety measures.

It is challenging to assess if some implications are overall welfare enhancing, especially when we consider that AI is a dual-use technology that creates externalities. The amount and kind of integration in the AI supply chain may be decisive in how effective different regulatory proposals for frontier AI models are. Antitrust policy will probably impact or complement regulation. Towards effective antitrust and regulatory intervention, there are major questions about the dynamics of the industry and its industry that need to be tackled.

## 6.1 Selected Research Questions

### 6.1.1 How might the prevailing market structure shape the trajectory of AI industry advancements?

Balancing safety and competition in the AI industry is a complex policy challenge, as these goals can sometimes be at odds. The effects of vertical integration on competition and R&D in the AI supply chain are unclear, requiring specific empirical research. While there are theoretical studies on arms races in the horizontal development of AI (e.g., Armstram, Bostrom, and Shalman, 2013), specific research on vertical relationships and empirical studies in either domain is lacking.

While increased R&D investment and faster technological progress are generally positive, the implications are ambiguous in the AI sector due to potential risks associated with advanced AI technologies. Vertical integration could encourage more investment in safety, especially if dominant firms prioritize it (Jensen, Emery-Xu, and Tragery, 2023). This could be beneficial for managing race dynamics among leading AI firms and fostering a safety-focused industry. However, less competition might lead to higher prices, reduced innovation, and fewer consumer choices, highlighting the need to carefully weigh these trade-offs.

### 6.2 How current market structure within the AI supply chain affect regulatory proposals?

Increased vertical integration in companies can make operational data less transparent, as integrated companies can closely control the flow of information. In the AI supply chain, this can include the number of AI accelerators purchased and the size of training runs. This opacity may complicate the task for external stakeholders like regulators and consumers in accurately monitoring and assessing the firm's activities that may be necessary to track for effective compute oversight. For example, information about Google's Tensor Processing Units (TPUs) is more restricted compared to Nvidia's Graphics Processing Units (GPUs). Increased vertical integration could, however, also mean that companies are more readily able to adhere to strict rules on data privacy and cybersecurity.

Having a generally more concentrated AI supply chain may also help to make coordinated efforts, such as, e.g., the MLCommons and Partnership for AI. Here, vertical integration could potentially provide integrated firms with the ability to establish and influence industry standards more easily. By controlling multiple stages of the supply chain, these companies can align their practices and technologies to create standards, which may be positive for AI governance concerns regarding standard-setting. In the context of compute oversight, this may be valuable to quickly adopt safety standards for AI accelerators, such as in-hardware monitors and shutdown mechanisms.

Conversely, it might increase the risk of collusive behavior, as observed in the DRAM market. This could stir the antitrust regulator’s concerns and this apprehension could inhibit potentially advantageous agreements between AI labs. Finally, more concentration could also mean a facilitated path for regulatory capture (see, e.g., Moshary and Slattery, 2023).

Drawing from the experiences of regulation development in industries such as electricity, civil aviation, and the banking sector can provide useful insights for shaping AI regulation. Lessons learned from these industries can help identify effective regulatory approaches, understand potential challenges, and inform the development of appropriate frameworks for the AI sector. Koesller and Schuett (2023) offer an overview of how risk assessments in other industries may impact this.

Insights from the literature regarding third-party reporting in supply chains can inform the development of AI regulations. This notion that vertically integrated firms are more capable of dodging taxation seems relatively established in the literature (see, e.g, De Paula and Scheinkman, 2010, Poremanz, 2015, Singh, 2020), but less so in regulation that demands the disclosure of key information. Lessons learned from public finance literature can contribute to the formulation of effective reporting and auditing mechanisms for AI development and deployment.

### 6.3 Will structural remedies be necessary to make effective regulatory frameworks in the AI industry?

As we discussed, the trade-offs between different forms of market integration — horizontal, vertical, and conglomerate — vary significantly. A key area for future research is identifying the optimal market structure for the AI supply chain that aligns with various policy objectives. For example, having a horizontally integrated market could potentially slow down AI development and reduce race dynamics at the potential expense of power concentration and less product diversity, while a less vertically integrated market might facilitate regulatory oversight.

Another issue to consider is whether unbundling principles, similar to those applied in the electrical and railway sectors, should be implemented in the frontier AI industry. The impact of such policies on consumer welfare is mixed and requires further examination. This is particularly relevant for enhancing third-party reporting mechanisms.

In terms of regulatory oversight, it would be beneficial to assess which types of integration hinder or support compute oversight. For instance, identifying clear bottlenecks where vertical integration should be avoided could enable more effective monitoring by regulators. Comparing incentive-compatible self-reports versus mandatory external audits can reveal which approach is more effective, depending on the level of integration in the industry. As Athey said “hopefully we will learn that we need to learn  more quickly this time than we have in the past” (Centre for Economic Policy Research, 2023)

# Appendices

## A. Mapping of the AI supply chain

| | |
|---|---|
| Overview - Mapping the AI supply Chain | A list of companies that have the potential to play a significant role, including relevant products, key personnel, market capitalization, and the industries in which they are active, among other factors. |
| Pairwise Relationship of Companies | A brief categorization and description of the pairwise relationships among all the companies covered in the mapping, including categories such as strategic partnerships, market customer-supplier relationships, and direct competition, among others. |
| Merger, Acquistions and Other Relevant Events | List of merger and acquisitions relevant for the AI industry as well other relevant events (e.g.: initial negotiation of acquisition that didn't progressed) |
| Antitrust litigations | Relevant antitrust cases for the AI industry in the USA, the EU and other relevant jurisdictions |
| Photolithography Companies | A brief description of this industry, including market shares. |
| Chip Fabricators | A brief description of this industry, including market shares. |
| AI Chip Designers | A brief description of this industry, including market shares. |
| AI Lab | A brief description of this industry, including market shares. |
| B2B Cloud | A brief description of this industry, including market shares. |

## B. Case study - OpenAI and Microsoft strategic partnership

### Introduction

In July 2019, OpenAI and Microsoft established a strategic partnership. In the deal, it is defined that OpenAI will exclusive license some of its frontier AI models to Microsoft and set Microsoft Azure as its exclusive cloud partner. In return, Microsoft invested USD 1 billion in OpenAI. In the announcement, OpenAI said that the company

> "is producing a sequence of increasingly powerful AI technologies, which requires a lot of capital for computational power. The most obvious way to cover costs is to build a product, but that would mean changing our focus. Instead, we intend to license some of our pre-AGI technologies, with Microsoft becoming our preferred partner for commercializing them."

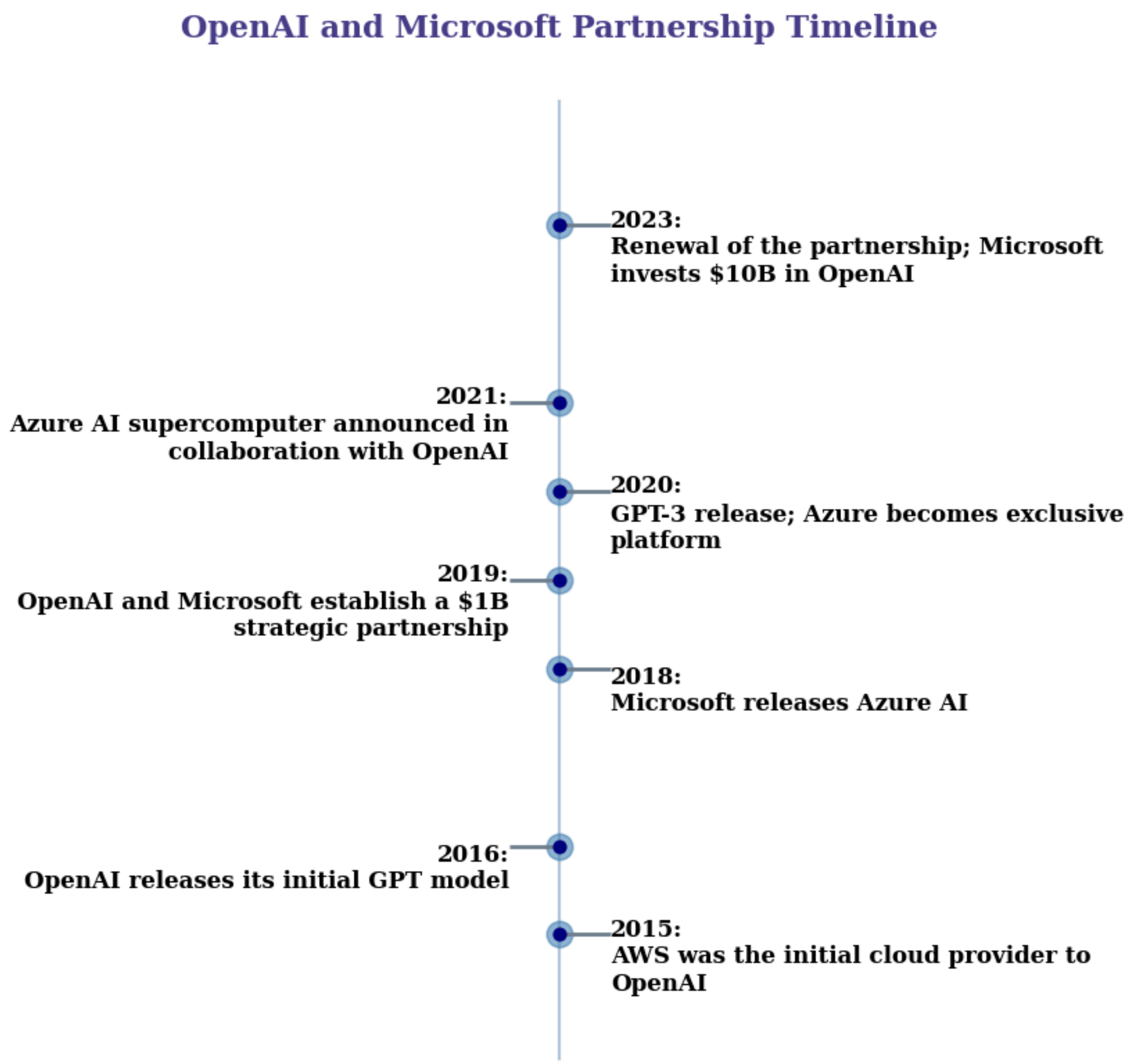

**Context and Negotiation Dynamics**

In the beginning of the company, Amazon Web Services was the cloud provider to the lab (OpenAI, 2015). Besides Azure, AWS and in-house solutions, Google Cloud would be the third contender for being the cloud provider to OpenAI.

It is unclear how exactly the negotiations were done. Google and OpenAI seemed to have a frigid relationship from the beginning. For instance, in the first interview after announcing the creation of OpenAI, to illustrate why the company was initially set as an nonprofit, Sam Altman said that "because we are not a for-profit company, like a Google, we can focus not on trying to enrich our shareholders [...] as time rolls on and we get closer to something that surpasses human intelligence, there is some question how much Google will share." Google acquisition of DeepMind in 2014 and its internal Google Brain team probably made Google not feel compelled to make a partnership with OpenAI. Reportedly, one of the goals of the OpenAI was to diminish market power that Google was establishing within the AI industry.

Microsoft, on the other hand, was actively seeking partners in AI to integrate it in its diverse portfolio of services. Microsoft vision has pushed them to seek partnerships with labs with the stated goal of developing AGI. Commenting on that, a Microsoft said that "Satya saw it coming and said 'let's do partnership with Open AI' and that mindset about how we can grow, be better all the time, brought us here" (Rikap, 2023)

Amazon's more application-focused approach to AI may have not combined that much with OpenAI's goals. "We are technology agnostic at Amazon. Other companies will go for the more expensive things. ChatGPT is an example", an Amazon employee said in an interview to Rikap (2023).

**Financial details**:

Microsoft and OpenAI renewed their strategic partnership in January 2023. As reported by Fortune (2023), the deal involved Microsoft investing 10 billion USD in OpenAI and getting 75% of OpenAI's profits until the investment was recovered. In summary, After that, Microsoft would get 49% of OpenAI's profits until it reached a total of 92 billion USD in earnings, at which point Microsoft's shares would revert back to OpenAI. In total, OpenAI needs to pay 105 billion USD to Microsoft to follow the deal. It is unclear if OpenAI receives royalties from Microsoft and how the 10 billion will be paid to OpenAI.

**Questions**

- How did OpenAI's previous relationship with AWS affect its decision to partner with Microsoft?
- Are the reported financial details accurate? Are there any conditions for OpenAI to receive royalties from Microsoft?
- Why did OpenAI choose to license some of its AI models exclusively to Microsoft?
- Why has OpenAI accepted paying 105 billion USD when valued at USD 28 BI (now reportedly 90)?
- Are there any exit clauses or contingencies in the partnership agreement?
- How much Microsoft is comfortable in depending upon foundation models developed by OpenAI?

## C. Case study - ASML's alliance with its customers to advance lithography

**ASML's Position:** Leading supplier in the semiconductor equipment industry, specializing in photolithography machines. Partnership with Intel, TSMC, Samsung. Synthetic buyback.

**Objective of the lliance**: Collaborate with key customers to advance lithography techniques and make the EUV technology commercially viable

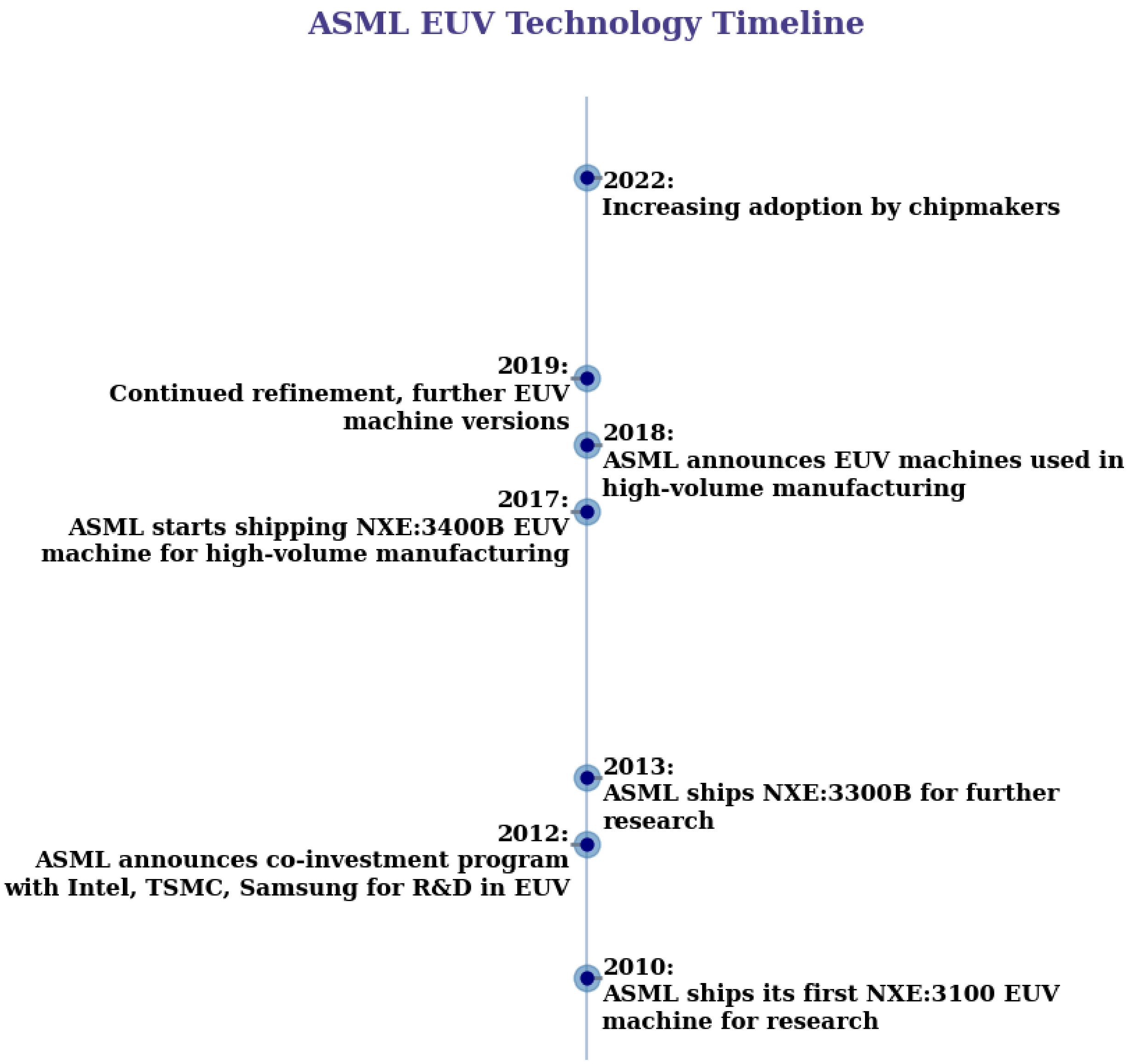

**Context and Negotiation Dynamics**:

- **Demand for progress**: Escalating need for smaller, more efficient semiconductor chips. Pressure to keep Moore's law
- **Technological ceiling**: Conventional lithography methods approaching their limitations. In 1998, industry experts converged to deem EUV as the next viable technology. By 2001, EUV LLC, a US consortium, had created functional EUV prototypes, but these lacked commercial viability. In a strategic move, ASML acquired rights and established a royalty agreement with EUV LLC.
- TwinScan and immersion technologies were early ASML wins and it was racing with Nikon and Canon to develop the new generation lithography technologies.
- **Mutual benefits**:
  - ASML receives direct feedback and insights about real-world chip manufacturing challenges.
  - Customers gain influence over the development of tools and early access to advanced technologies.

**Financial Details:**

- **Investment in R&D**: Significant funds allocated for the development of Extreme Ultraviolet (EUV) lithography and related technologies.
  - **Intel**: In July 2012, Intel agreed to purchase a roughly 15% stake in ASML for about €2.5 billion. Intel also committed an additional €830 million to fund ASML's research and development efforts over five years.
  - **TSMC**: Later in July 2012, TSMC announced that it would acquire a 5% stake in ASML for €838 million and invest an additional €276 million in R&D over five years.
  - **Samsung**: In August 2012, Samsung joined the program, acquiring a 3% stake in ASML for €503 million and committing another €276 million for R&D over the next five year
- **Potential returns**: Enhanced precision in chip design promises greater efficiency, potentially leading to higher profit margins for both ASML and its customers. Intel, TSMC and Samsung were eager to chips with more precision

- **Risk factor:** High costs and uncertainties associated with pioneering and implementing new technologies.

Key Takeaways:

- **Technological leap**: The development and rollout of EUV lithography solved the problem where upstream companies didn't want to invest in R&D because the benefits mainly went downstream. This was addressed through a customer co-investment program.
- **Strengthened industry relations**: ASML's alliance model demonstrates the benefits of collaboration between manufacturers and its clients. ASML customers in the alliance gain early access to new technologies and can influence their development, potentially gaining a competitive advantage in the semiconductor market.
- **Affirmation of industrial dominance**: While Canon and Nikon have not succeeded in developing EUV technologies—with Canon not leading in past generation immersion technology and Nikon ceasing EUV efforts since 2011—they continue to be significant players in other less advanced lithography technologies.

**Questions**

- What conditions led Intel to invest a more significant amount in ASML? Which advantages did they negotiate? Why didn't they demand exclusivity?
- What are the specifics of ASML's royalty agreement with EUV LLC?
- Has something specific made antitrust?
- How does the alliance handle intellectual property rights?
- Nikon has really given up trying to develop EUV? Is there any other realistic contender?
- How did they manage risks in the partnership? Were there exit clauses?
- How other technologies may be relevant?
- To what extent were Moore's laws sustained by economic pressures? To what extent was the government? Search for sources, how much government contracts were relevant, etc.

## D. Case study - Nvidia and ARM talks

- Entities involved: Nvidia; ARM; FTC;
- Objective of the talks: Nvidia's intention to acquire ARM in a bid to expand its influence and consolidate its stance in the semiconductor domain.;

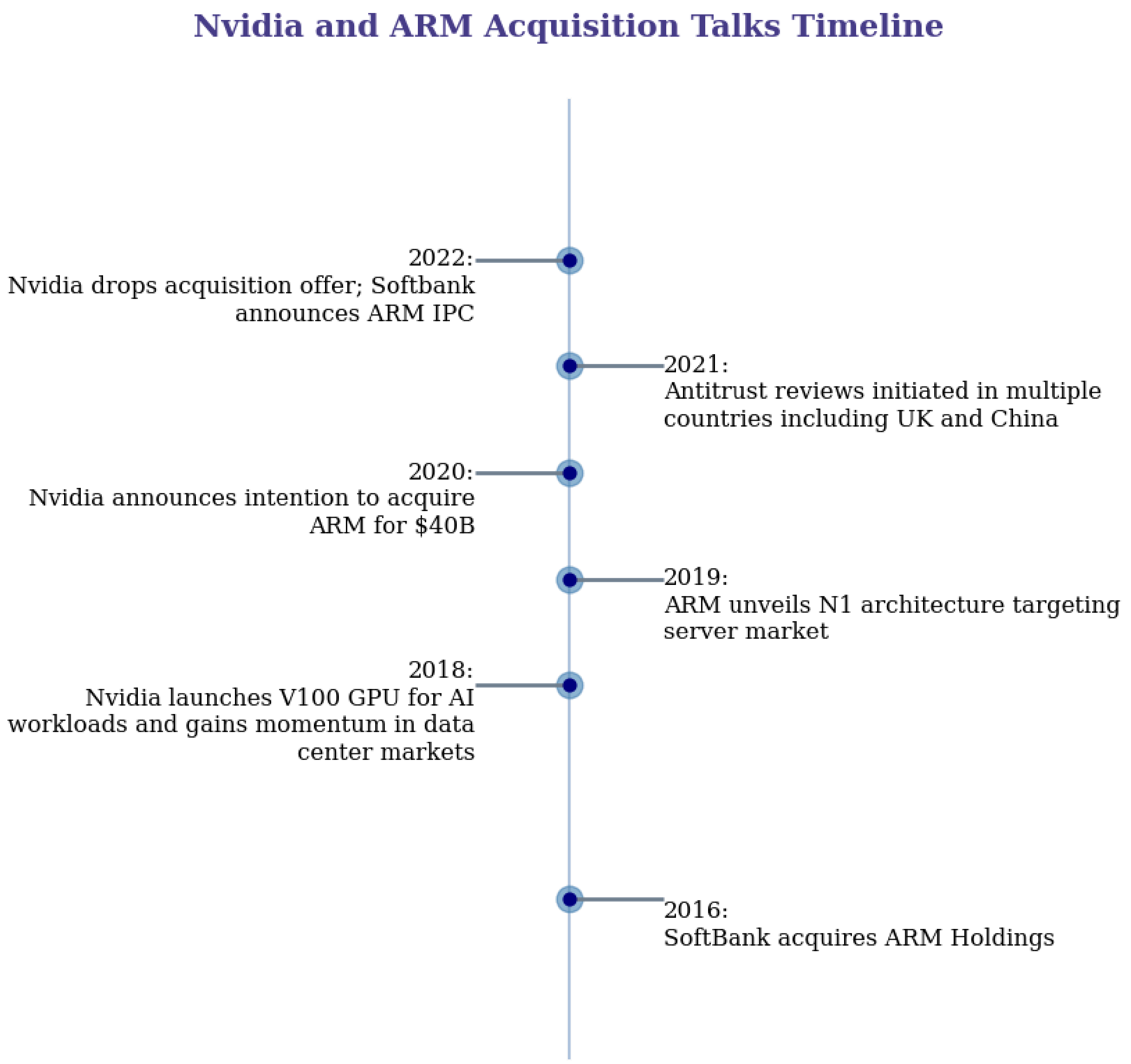

**Context and Negotiation Dynamics:**

- Strategic importance: As NVIDIA develops frontier GPUs and ARM develops the most used core IP architecture, the proposed merger would be of great importance to the semiconductor industry. They could potentially have enjoyed synergies between their activities.

- FTC's concerns: The FTC raised flags about Nvidia potentially obtaining an unfair competitive advantage.. They believed Nvidia might restrict ARM's Core IP licensing to other chip designers, potentially leading to a monopoly.
- After the termination of the acquisition attempt by Nvidia, the FTC said through a press release that "*The termination of what would have been the largest semiconductor chip merger will preserve competition for key technologies and safeguard future innovation. This result is particularly significant because it represents the first abandonment of a litigated vertical merger in many years." (2022)*
- After Nvidia dropped the offer, Softbank announced the intent of doing an IPO of ARM.

**Financial Details:**

- Deal value: Nvidia's intended acquisition was priced at around $40 billion.
- Payment structure: A proposed mix of Nvidia shares and cash considerations.

**Key Takeaways:**

- As highlighted by FTC, this was the first termination of a litigation vertical integration in a significant time
- Shift in competitive dynamics: Had the acquisition been successful, it could have dramatically altered the competitive dynamics in the semiconductor sector.

**Questions**

1. What strategic advantages did Nvidia aim to gain through the acquisition of ARM? What specific synergies were expected between Nvidia's GPUs and ARM's Core IP?
2. How might the acquisition have impacted ARM's existing licensing agreements with other chip designers?
3. What were the alternative strategies considered by Nvidia after the FTC's concerns? What contingency plans did Nvidia have in place?
4. Were there any discussions or preparations around antitrust compliance prior to the FTC's intervention?